\documentclass[twocolumn]{aastex631}

\newcommand{\eqref}[1]{ (\ref{#1}) }
\def\Ra{\mbox{\rm Ra}}

\def\Ro{\mbox{\rm Ro}}

\def\Pra{\mbox{\rm Pr}}

\def\Re{{\rm Re}}

\def\apjl{ApJL}
\def\apjs{ApJ}
\def\aap{A\&A}

\def\Rs{R_{\odot}}
\def\prot{P_{\rm rot}}

\def\tauc{\tau_{\rm c}}

\newcommand{\brac}[1]{\langle #1 \rangle}
\newcommand{\mean}[1]{\overline{#1}}

\usepackage{amssymb}
\usepackage{xcolor}
\usepackage{bm}
\usepackage{enumerate}
\usepackage{hyperref}
\usepackage{float}
\usepackage[utf8]{inputenc}
\usepackage{graphicx}
\usepackage{graphics}
\graphicspath{{figs/}{png/}}
\submitjournal{ApJ}

\shorttitle{Grid-resolution dependency on ILES of solar convection}
\shortauthors{Guerrero et al.}
\graphicspath{{figs/}{png/}}

\begin{document}

\title{Implicit large eddy simulations of global solar convection: effects of
numerical resolution in non-rotating and rotating cases}

\correspondingauthor{G. Guerrero}
\email{guerrero@fisica.ufmg.br}

\author[0000-0002-2671-8796]{G. Guerrero}
\affiliation{Physics Department, Universidade Federal de Minas Gerais \\
Av. Antonio Carlos, 6627, Belo Horizonte, MG 31270-901, Brazil}
\affiliation{New Jersey Institute of technology, Newark, NJ 07103,USA}

\author[0000-0001-7483-3257]{A.~M. Stejko}
\affiliation{New Jersey Institute of technology, Newark, NJ 07103,USA}

\author[0000-0003-0364-4883]{A.~G. Kosovichev}
\affiliation{New Jersey Institute of technology, Newark, NJ 07103,USA}

\author[0000-0001-7077-3285]{P.~K. Smolarkiewicz}
\affiliation{National Center for Atmospheric Research, Boulder, Colorado, USA}

\author[0000-0002-9630-6463]{A. Strugarek}
\affiliation{AIM, CEA, CNRS, Université Paris-Saclay, Université Paris Diderot, Sorbonne Paris Cité, France}

\begin{abstract}
  Simulating deep solar convection and its coupled mean-field motions is a formidable
  challenge where few observational results constrain models that suffer from
  the non-physical influence of the grid resolution. We present hydrodynamic
  global Implicit Large-Eddy simulations (ILES) of deep solar convection performed
  with the EULAG-MHD code, and explore the effects of grid resolution on
  the properties of rotating and non-rotating convection.  The results,
  based on low-order moments and turbulent spectra reveal that convergence could
  be achieved in non-rotating simulations provided sufficient resolution in
  the radial direction. The flow is highly anisotropic, with the
  energy contained in horizontal divergent motions exceeding by more than three
  orders of magnitude  their radial counterpart. By contrast, in rotating simulations
  the largest energy is in the toroidal part of the horizontal motions.  As the
  grid resolution increases, the turbulent correlations change in such a way that
  a solar-like differential rotation, obtained in the simulation with the coarser
  grid, transitions to the anti-solar differential rotation.
  The reason for this change is the contribution of the effective viscosity
  to the balance of the forces driving large-scale flows. As the effective viscosity
  decreases, the angular momentum balance improves, yet the force balance in the
  meridional direction lessens, favoring a strong meridional flow that advects angular
  momentum towards the poles. The results suggest that obtaining the correct
  distribution of angular momentum may not be a mere issue of numerical resolution.
  Accounting for additional physics, such as magnetism or the near-surface shear
  layer, may be necessary in simulating the solar interior.
\end{abstract}


\section{Introduction} 
\label{sec:intro}

Turbulent convection is observed at the solar photosphere in the form of granulation, 
with scales of several thousands of kilometers, and supergranulation, with scales of 
$\sim 30000$ km. From theoretical considerations it is expected that larger, 
self-similar, motions exist at deeper layers. Their characteristics, however, 
remain evasive to observations.
The properties of solar convection at deeper layers are relevant for 
understanding the physical processes that drive and shape 
the large-scale flows observed in the Sun, namely the differential rotation (DR) 
and the  meridional circulation (MC). 
Furthermore, they may be instrumental for the solar dynamo process. 

The knowledge that we have about the deep turbulent motions is based on the 
one-dimensional mixing-length theory \citep[MLT][]{bohm-vitense58,kippenhahn}. 
The results of MLT models predict that spatial and temporal scales progressively 
increase with depth. Nevertheless,  observations of turbulent motions just beneath the 
supergranulation layer offered conflicting results such that the MLT scenario 
cannot be confirmed. From one side, \citet{hanasoge+12} found that the 
larger scales have convective power several orders of magnitude smaller than the MLT 
prediction. On the other hand,  the results of \citet{greer+15} suggested 
spectral energies more in line with the theory, yet perhaps with larger power
than expected. This ambiguity has often been called the solar convective 
conundrum. 

\citet{proxauf_phdT} and collaborators have recently identified several
incompatibilities in the comparison of the previous results,
proposed the required corrections, and performed new inferences. Despite
modifications, the disparity between the two helioseismic inferences 
still spans three to four orders of magnitude. Their new surface granulation 
tracking, and ring-diagram analysis measurements indicate that at low angular
harmonic degree, $\ell$, the trend of the power spectra agrees with the corrected 
results of \citet{hanasoge+12}, with slightly higher power (about one order
of magnitude). Other analysis seem to confirm the existence 
of structures larger than supergranulation but with rather small power 
\citep{Hathaway+21,getling+22}. At deeper layers, the properties of 
convection are still a mystery,
and we rely on global simulations of turbulent rotating convection to 
get access at least to some of its characteristics.  
Unfortunately, simulations of solar convection are not exempt from difficulties. 

The big goal of global convection simulations, precluding so far 
the possible development of magnetic fields,  is developing  motions able to carry out 
most of the solar luminosity, $L_{\odot}$, simultaneously producing a mean angular 
velocity in agreement with the helioseismic observations \citep{schou+etal_98}, and
a meridional circulation with poleward migration at surface levels 
\citep[e.g.,][]{UL10,HU14}. The meridional circulation  profile in the
deep convection zone is still 
under investigation and a matter of debate \citep[see][for inferences with two and 
one meridional cells per hemisphere]{chen+17,gizon+20,steyko+21}.

The parameter regime of the solar interior where convection must transport
energy and angular momentum is characterized by the Rayleigh ($\Ra$) number 
${\cal O} (10^{20})$, the Reynolds number ($\Re$) ${\cal O} (10^{12})$, and the 
Prandtl ($\Pra$) number ${\cal O} (10^{-7})$ \citep[see Table 2 in][]{ossendrijver03}.  
The separation of the dynamical scales
determined by these parameters requires an enormous resources to be  resolved
by direct numerical simulations (DNS). Modern supercomputers
are still far from allowing DNS at such resolution. Thus, for the current DNS, the 
dissipative coefficients that define the non-dimensional numbers are orders
of magnitude larger than the microscopic values of viscosity and/or thermal 
conductivity. Considering that the values of these coefficients are consistent
with the estimated turbulent dissipation, these simulations can be viewed as
large-eddy simulations (LES) which preclude non-dissipative contribution of 
the unresolved scales.

In state-of-the-art simulations in which the radiative flux driving convection at 
the bottom of the convection zone corresponds to $L_{\odot} = 3.83\times10^{26}$ W,  
and the rotation 
imposed on the frame is the solar sideral rotation, 
$\Omega_{\odot} = 2.97 \times 10^6$ s$^{-1}$, the convective 
heat flux results in enhanced turbulent velocities whose energy spectrum disagrees
with the helioseismic observations described above \citep{gizon+12}.  Additionally, 
higher velocities imply shorter convective correlation times, $\tauc$, and, therefore,
larger Rossby numbers,  which is the ratio between rotation and convection
timescales, $\Ro = \prot/\tauc$. This ratio is an indication of how much 
the motions are constrained by the Coriolis force.  It is a fairly robust result of 
global convective simulations \citep{KMGBC11,GSKM13b,GYMRW14,FM15} that as 
the motions get less rotationally constrained (high $\Ro$), the differential 
rotation profile 
results in faster poles and slow equator, at odds with the solar-like profile. 

A solar-like differential rotation  profile with an accelerated equator 
may be recovered by increasing the rotation rate by a factor of two or 
three \citep{BBBMT08, hotta18}.  Another alternative
is diminishing the convective transport by increasing the 
radiative diffusion coefficient \citep{miesch+08,kapyla+14,FM15}, or 
by decreasing the luminosity of the simulation \citep{GSKM13b,hotta+15}. 
The latter options artificially decrease the strength of the convective 
flows.  In general, these alternatives decrease the Rossby number. 

Large-scale or small-scale 
magnetic fields may also help to reproduce  the solar 
differential rotation.  They contribute in two different ways: by modulating
the convective heat transport \citep{fan+14,KKBOP15,guerrero2019sets} which 
decreases the turbulent 
velocities (lower $\Ro$), and through the direct transport of angular momentum 
via the large and small scale Maxwell stresses.  The contribution of the 
large-scale magnetic field has been verified by lower resolution simulations,
which allow for a long temporal evolution and, therefore, the excitation 
and sustainment of the large-scale dynamo.  The contribution of the small-scale 
dynamo has been recently
identified in the high-resolution simulations by \citet{hotta+21}. They suggested
that the small-scale magnetic field diminishes the convective power and that
a meridional flow, developed to balance the Maxwell stresses, transports
angular momentum towards the equator.  However, 
these simulations have run only for short periods of time, such that the 
large-scale magnetic field did not develop. Thus, there is still no clarity 
on the distribution of small-scale magnetic fields in the presence of 
large-scale dynamo, and it is uncertain what is the role of both contributions 
on the angular momentum transport. 

Most simulations described above suffer from the contribution of 
unrealistically large dissipative coefficients caused by the limited
spatial resolution.  Large viscous dissipation, for 
instance, is needed to guarantee numerical stability.  It is combined with the
numerical dissipation associated with the numerical schemes.  Both 
contributions directly affect the transport of linear and angular momentum,
whereas in the Sun, the microscopic viscosity is negligible. 
Therefore, it is desirable albeit challenging to achieve converged solutions
independent on numerical resolution.
In other words, it is appealing having a system where the 
dissipation coefficients are insignificant 
and the dynamics is governed by well-resolved turbulent motions
subject to well-defined boundary conditions.  However, as discussed above,  it does 
not seem to be the case for the current models of deep solar convection. 
As a matter of fact, very few works have systematically explored the
role of resolution  on the properties of convection.
The work of \citet{featherstone+16a} is one exception, for 
non-rotating convection in the sphere, with remarkable results.
They are able to achieve simulations where the amplitude of the kinetic energy 
becomes independent of the $\Ra$ values, yet the distribution of root mean square (RMS) 
velocities and 
kinetic turbulent spectra continue changing as $\Ra$ further increases. 

Large-eddy-simulations including explicit or implicit 
\citep{Grinstein+07} turbulent sub-grid scale (SGS) contribution may be a 
computationally less expensive alternative to DNS of solar (stellar) convection and 
dynamo. LES have been routinely and successfully used in engineering and 
meteorological problems. This approach has also been used for modeling local 
convection and small-scale dynamo \citep{wray+15,rogachevskii+11,kitiashvili+15}.  
Nevertheless, it has not been broadly considered in the problem of global 
simulations of solar and stellar convection and dynamo. This
is understandable since, unlike laboratory experiments, or weather and climate
studies, there is not sufficient available data nor convergent DNS results of the 
same problem.  This prevents the comparison and characterization of the LES
results and the SGS models. 

The implicit LES simulations performed with the EULAG-MHD code, based in 
a non-oscillatory forward-in-time algorithm,  have been successful in reproducing
at lower resolution some characteristics of the solar activity cycle \citep{GCS10}, 
and have also been used to study stellar activity \citep{SBCBdN17,guerrero2019sets}.
In the code, the dynamical equations may be solved in their inviscid form,  
with the viscous contribution given by the minimized truncation terms,
allowing to achieve the higher possible level of turbulence, i.e., a larger 
value of $\Re$, for any resolution.  In this form, it is less computationally expensive 
to search for solutions to the problem independent of the non-physical 
influence of numerical resolution.
We have demonstrated convergence in the solutions
of simulations of 2D convection for a range of different mesh sizes
\citet{nogueira+22}. The low-order moments in the physical space as well as 
the turbulent spectra of
kinetic and thermal energy were used as criteria of convergence.  We attribute 
the convergence of the results in the low-resolution cases to the implicit 
SGS viscosity of the non-oscillatory discrete advection method,
which allows the 
back-scatter of energy from small to large scales and dissipation at the 
minimal scale, close to the limit of numerical resolution. 
For progressively larger resolutions, however,  large 
and small scale motions coexist in a system dominated by a mesh-independent 
turbulent dissipation. In these cases the contribution from the small scales 
seems to be resolved.

In this paper, we continue the study presented in \citet{nogueira+22} while
approximating our numerical model to the solar interior by considering spherical
geometry.  Additionally, we consider the effects of rotation on the turbulent 
motions and explore the development of mean flows in models where all parameters
are kept constant except the grid resolution. This problem is more complicated
than the Cartesian case. On the one hand, the spherical geometry imposes curvature 
effects and numerical stiffness in polar regions.  On the other hand, gravity and 
rotation result in strongly anisotropic convective motions upon which the amount of 
viscous resistance, wherever is its origin,  turns out to be highly relevant.  
We anticipate that for non-rotating cases,  hints of convergence are
achieved at a resolution 
consistent with the Cartesian cases. For the rotating cases, the resulting large-scale 
motions are substantially dependent on the resolution due to the decreasing 
contribution of effective viscosity to the net transport of angular momentum.
We characterize these changes through a comparative analysis 
of the spectral properties of the fluid, and the balance of the azimuthal 
and meridional forces that drive these large-scale motions. To differentiate 
the most robust features of global convection from those that seem to be 
model dependent,  we extend our comparison to previous findings of similar 
models, and to solar observations, when possible.
Even though we do not achieve convergence, 
the analysis contributes to understanding the sustainment of solar mean-flows 
and allows us to hypothesize possible ways to achieve solutions to the 
convective conundrum. 

This paper is organized as follows. In Sect.~\ref{sec:model}, we describe
the numerical model. The results of non-rotating and rotating cases are 
presented in Sect.~\ref{sec:results}, and in Sect.~\ref{sec:conclusions}, we
discuss our findings and expose the conclusions of this work.

\section{Numerical model}
\label{sec:model}

The EULAG-MHD code  \citep{SC13} ---a specialized variant of the 
original EULAG code \citep{prusa2008eulag}--- is used  to perform global 
anelastic convection  simulations. EULAG-MHD is based on the multidimensional 
positive-definite 
advection transport algorithm \citep[MPDATA;][]{smolarkiewicz2006multidimensional}.
It is a non-oscillatory forward-in-time advection solver with second-order accuracy in
space and time.  
The code allows simulations to be run as ILES
without explicit dissipation, yet it also may be used for DNS with 
explicit dissipation. 
The domain corresponds to a global spherical shell  
with the radial coordinate, $r$, 
covering the upper fraction of the radiative zone, from $r_b=0.6\Rs$,  and the 
convection zone (CZ) up to $r_t = 0.96\Rs$. We exclude the upper layers of the
CZ where compressibility and radiative transfer play a major role in the dynamics 
of convection. 

The code integrates the following set of Navier-Stokes 
equations governing mass, momentum, and energy conservation,
\begin{equation}
    \nabla\cdot\rho_r\textbf{u}=0,
    \label{eq_mass_cons_nc}
\end{equation}
\begin{equation}
	\frac{d\textbf{u}}{dt} + 2 {\bf \Omega} \times {\bf u} =
    - {\nabla \pi^{\prime}}
    - {\textbf{g}\frac{\Theta^{\prime}}{\Theta_r}} ,
    \label{eq_momentum_nc}
\end{equation}
\begin{equation}
    \frac{d\Theta^{\prime}}{dt} = -\textbf{u}\cdot\nabla\Theta_a - \alpha \Theta^{\prime}
    \label{eq_energy_nc}
\end{equation}

\noindent where $d/dt = \partial/\partial{t} + \textbf{u}\cdot\nabla$, $\textbf{u}$ is 
the velocity field in a frame rotating with
angular velocity 
${\bf \Omega} = (\Omega_r, \Omega_{\theta}, \Omega_{\phi})=\Omega_0(\cos\theta,
-\sin\theta,0)$, $\rho_r$ is the reference state density, 
which in the anelastic approximation is a function of radius
only \citep{LH82}; $\pi^{\prime}$  
is the density normalized pressure perturbation, $p^{\prime}/\rho_{r}$; 
$\textbf{g}= -g {\bf \hat{e}}_r$ is the gravity acceleration adjusted to fit the
solar gravity profile, and $\Theta$ is 
the potential temperature defined as $\Theta=T\left(P_b/P\right)^{R/c_p}$, where $T$ is the 
temperature, $P$ is the pressure, 
$P_b$ is the pressure at the bottom of the domain,
$R = 13732$ J K$^{-1}$ kg$^{-1}$ is the universal gas constant for a 
monoatomic hydrogen gas, and $c_p=2.5 R$, is the specific heat 
at constant pressure. The potential temperature is equivalent to the 
specific entropy through the 
relation $ds = c_{p}d(\ln \Theta)$. The subscripts $r$ and  $a$ refer to the 
reference and ambient states, and 
the superscript $\prime$ means perturbations of a quantity about the ambient profile. 
Perturbations of $\Theta$ are related to perturbations of temperature by 
the anelastic approximation,  
$T^{\prime} = \Theta^{\prime} T_{a} / \Theta_{a}$. 
The energy equation contains a term forcing the adiabatic perturbations about the 
ambient state and a thermal relaxation term that damps these perturbations in an
inverse time scale, $\alpha=1/\tau$ \citep[cf.][for a discussion]{cossette2017magnetically}.

\begin{figure}%
    \begin{center}
    \includegraphics[width=\columnwidth]{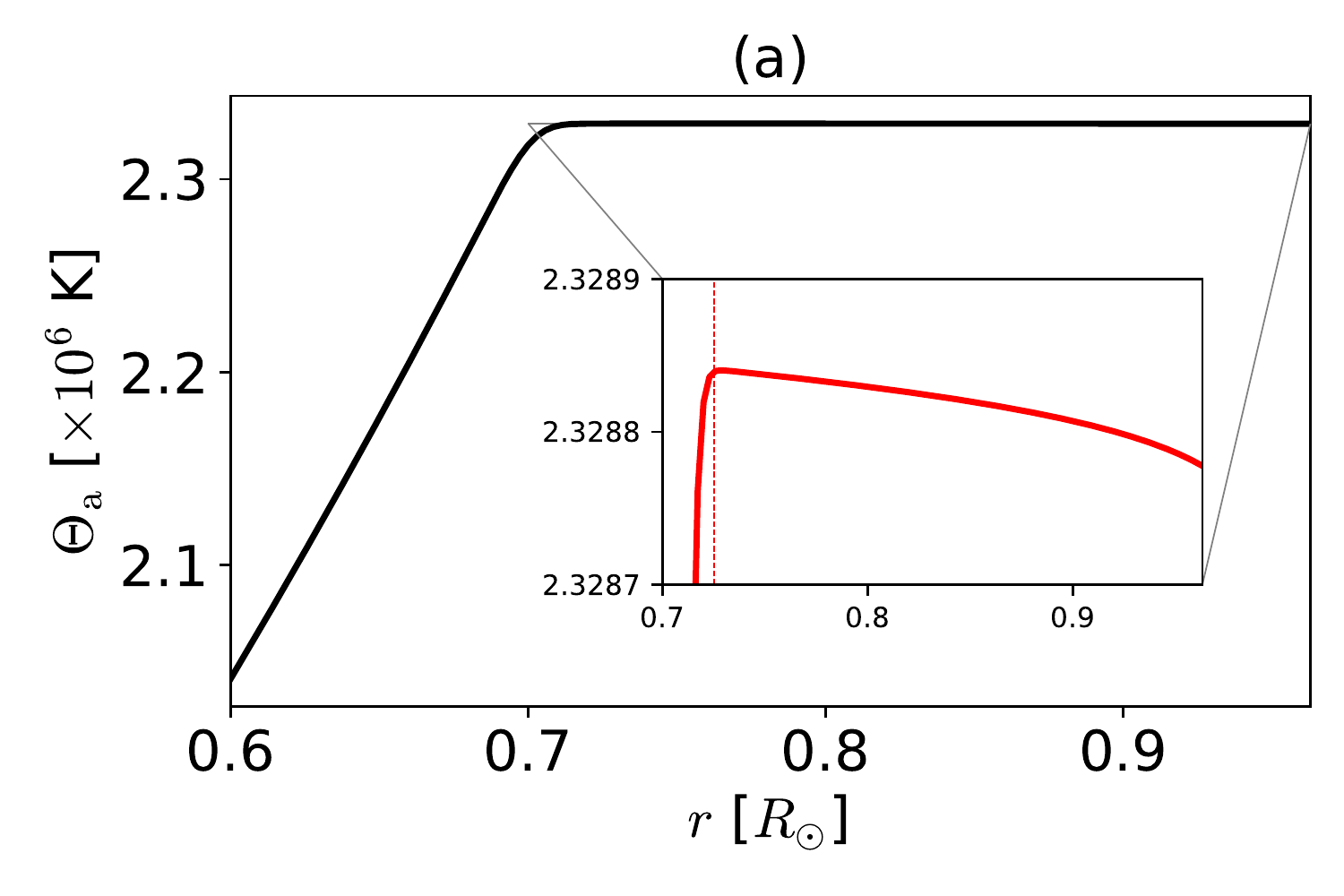} \\
    \includegraphics[width=\columnwidth]{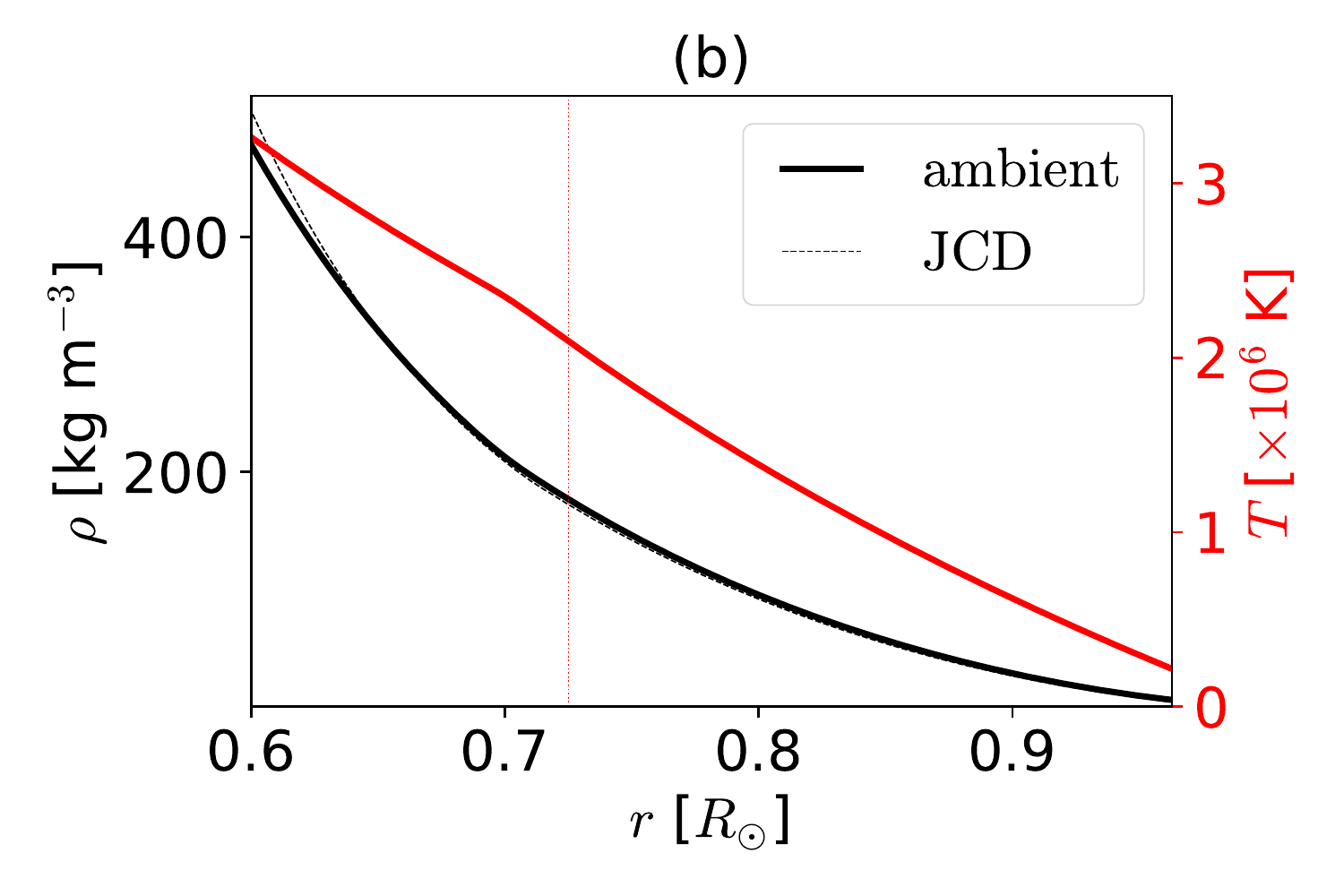} 
    \caption{(a) Radial profile of the ambient potential temperature, $\Theta_a$.
    The insert shows a close-up for $0.7 < r/\Rs < 0.96$ depicting a negative
    radial gradient of $\Theta_a$. The difference between the bottom and top
    of the unstable layer is $62$ K. (b) 
    Radial profiles of the ambient density (black line) and temperature 
    (red line). The dotted lines correspond to the solar structure 
    model of \citet{CD+96}. From the bottom to the top of the domain there
    are $\sim4.5$ density scale heights. From the base of the
    convection zone, $r > 0.7\Rs$ there are $\sim3.7$ density scale heights.
    The vertical red lines show the bottom of the unstable layer located at 
    $r=0.725 \Rs$.}%
    \label{fig_ambient_state}%
    \end{center}
\end{figure}

The ambient state defining the thermodynamic variables, $\rho_a$, $\Theta_a$, 
and $T_a$ in equations~\eqref{eq_mass_cons_nc}-\eqref{eq_energy_nc} is a 
particular solution of the  hydrodynamics equations. In this work, the  
ambient state considering hydrostatic equilibrium for a 
non-rotating atmosphere is constructed by solving the following equations,
\begin{equation}
    \frac{\partial T_a}{\partial r} = 
    -\frac{g}{R(m+1)},
    \label{est_amb1}
\end{equation}
\begin{equation}
    \frac{\partial\rho_a}{\partial r} = 
    -\frac{\rho}{T_a}
    \left(\frac{g}{R} + \frac{\partial T_a}{\partial r} \right),
    \label{est_amb2}
\end{equation}

\noindent where $m=m(r)$ is the polytropic index.  Solutions of 
equations~\eqref{est_amb1} and \eqref{est_amb2} with  $m\geq 1.5$ correspond to stable 
stratification, while solutions for $m<1.5$ correspond to convectively unstable states.

The ambient state with a stable layer at the bottom of the domain is built by
setting $m_{s}=2.5$ for $r\le0.7\Rs$, and a marginally unstable convection zone 
with $m_{u}=1.49997$ for $r > 0.7\Rs$.  This is achieved by considering a radial 
profile of the polytropic index, 

\begin{equation}
m(r) = m_{s} - \frac{1}{2}(m_{s} - m_{u}) \left[1 + \text{erf}\left(\frac{r-r_{1}}{w}\right) \right] \;,
\label{eq_m}
\end{equation}

\noindent where the transition between zones of different $m$ is made through the 
erf functions with $r_1 = 0.7\Rs$ and $w=0.01 \Rs$. Equations~\eqref{est_amb1} 
and \eqref{est_amb2} are integrated numerically with $\rho_{r_1}=208$ kg/m$^3$ and
$T_{r_1} = 2.322 \times 10^{6}$ K at the interface between the stable and the 
unstable layers,
The pressure is computed via the ideal 
gas equation of state, $P_a=R \rho_a T_a$. The resulting profile of $\Theta_a$ 
as a function of radius is shown in Figure~\ref{fig_ambient_state}(a). 
In the convective zone, the slope of $\Theta_a$ is slightly 
negative with respect to the $r$ coordinate as it can be seen in the figure insert. 
The negative slope of $\Theta_{a}$ ensures that this zone is unstable to convection, with 
the difference of $\Theta_{a}$ between the bottom and top of the convectively unstable 
layer being $62$ K. The reference potential temperature $\Theta_r = T_{r_1}$.
Finally, for all the simulations, $\alpha = 1/\tau = 1.29\times10^{-8}$ s$^{-1}$
is considered. Fig.~\ref{fig_ambient_state}(b) shows the radial profiles of the 
density, $\rho_a$, and the temperature, $T_a$. The entire radial domain 
encompasses $4.5$ density scale heights with $3.7$ density scale heights 
corresponding to the unstable layer.

The boundary conditions for this setup are: impermeable,  stress-free 
conditions for the velocity field at the two radial ends of the domain.
Null convective radial flux is considered as thermal boundaries at the bottom
and top, as it has been used in previous works in the literature 
\citep[e.g.,][]{fan+03,hotta+15}.  For most of the cases the initial 
conditions are white noise perturbations in  
$\Theta^{\prime}$,  introduced only in the unstable layer and with 
maximum amplitude $0.1$ K.

\begin{deluxetable*}{ccccccccc}
\tablenum{1}
\tablecaption{Simulation parameters and results \label{table:1}}
\tablewidth{0.9\columnwidth}
\tablehead{
\colhead{Simulation} & \colhead{$N_{\phi},N_{\theta}, N_{r}$} & 
	\colhead{$dt$ [s]} & 
  \colhead{Runtime [years]} & 
  \colhead{$\brac{ U_{\rm rms}} $  [m/s]} & 
	\colhead{$\brac{ \Theta^{\prime}} $ [K]} & 
  \colhead{$\brac{ L_{\rm e}}$ [$L_{\odot}]$} &
	\colhead{$\Ro$} 
%	\colhead{$\nut$ [$\times 10^9$ m$^2$/s ]} 
}
\startdata
R1    &   128,64,64     & 1400 & 200 & 54.07   & 21.49 & 0.38 & -    \\       %& 3.14   \\
R2    &   256,128,128   & 500  &  80 & 51.70   & 13.78 & 0.25 & -    \\       %& 3.00   \\
R4    &   512,256,256   & 200  &  26 & 61.92   &  9.45 & 0.17 & -    \\       %& 3.59   \\
R8    &   1024,512,512  & 50   & 2.8 & 69.24   &  9.23 & 0.14 & -    \\       %& 3.98   \\
R1x   &   128,64,256    & 800  &  40 & 65.15   &  9.37 & 0.16 & -    \\\hline %& 3.76   \\\hline
R1x24 &   128,64,256    & 800  &  80 & 45.85   &  9.50 & 1.30 & 0.56 \\       %& 2.68   \\
R2x24 &   256,128,256   & 600  &  60 & 46.90   &  9.60 & 1.37 & 0.58 \\       %& 2.72   \\
R4x24&   512,256,256    & 200  &  20 & 52.32   &  9.54 & 1.34 & 0.65 \\       %& 3.03   \\
\enddata
\tablecomments{Parameters and results of the simulations presented in
  this work. The anglular brackets, $\brac{}$, correspond to volume and
  temporal averages considering only the convection zone, i.e., an
  average over radius of the profiles presented in Figs.~\ref{fig:prof_r},
  \ref{fig:cflux}, \ref{fig:prof_rot_r}, and \ref{fig:enth_flux_rot}. 
	The Rossby number is defined as, $\Ro = \prot/\tauc$, where, 
	$\tauc = \ell_c/\brac{U}_{\rm rms}$, is the convective turnover time. For
	simplicity, the convective correlation length,  $\ell_c$, is defined as
	the thickness of the unstable layer. 	}
\end{deluxetable*}

\section{Results}
\label{sec:results}

In the simulations discussed below, the ambient stratification
and the thermal relaxation time scale are kept constant, therefore, the
forcing and the thermal relaxation terms in Eq.~(\ref{eq_energy_nc}) are 
theoretically the same in all models (i.e., in the hypothetical limit of 
the converged numerical solutions).
Non-rotating models ($\Omega_0=0$) with five 
different numerical resolutions (see Table~\ref{table:1}) are presented. Based on the 
results of these models three rotating simulations, with $\Omega_0=3.03\times10^{-6}$ 
s$^{-1}$ ($\prot=24$ days) for three different resolutions are performed. 
For this rotation rate, the low-resolution case 
produces a solar-like differential rotation profile. 
The runtime of each 
simulations, here considered as the physical time, in years, that the variables 
are  evolved, is presented in Table~\ref{table:1}. 
The simulation R8 is started from a remeshed snapshot of the simulation 
R4 in the relaxed state and ran for $\sim 2.8$ Earth years.  The temporal averages 
are calculated from outputs with a cadence of 1 month over a time range of 
5 years. For simulation R8 only 6 months of simulated data were considered.

\subsection{Non-rotating convection}
\label{sec:nr}

\begin{figure*}%
    \begin{center}
    \includegraphics[width=0.27\textwidth]{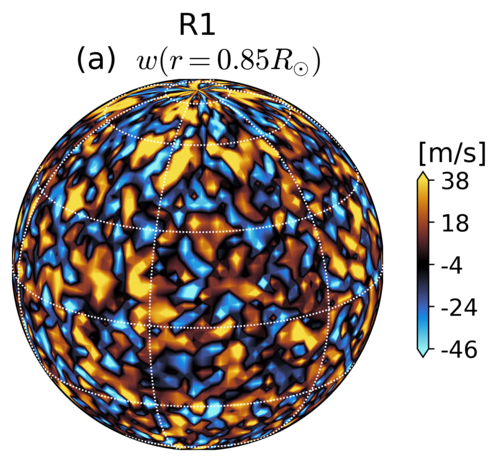} 
    \includegraphics[width=0.17\textwidth]{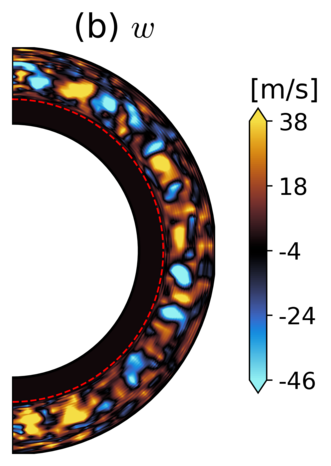}
    \includegraphics[width=0.17\textwidth]{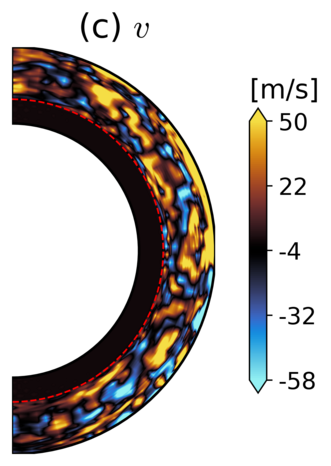} 
    \includegraphics[width=0.17\textwidth]{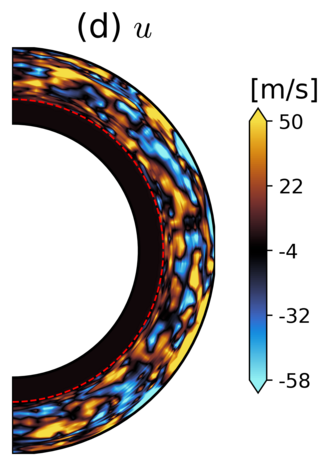} \\
    \includegraphics[width=0.27\textwidth]{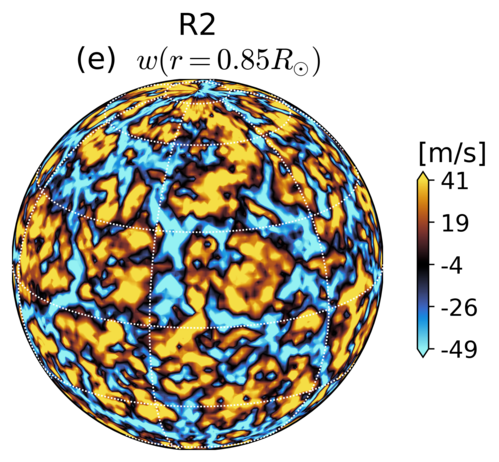} 
    \includegraphics[width=0.17\textwidth]{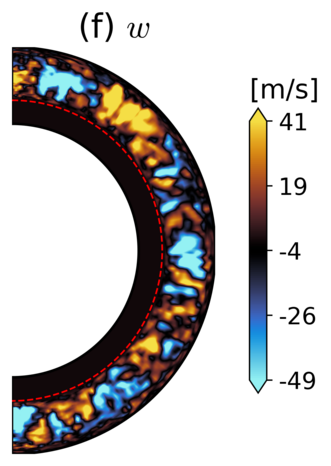}
    \includegraphics[width=0.17\textwidth]{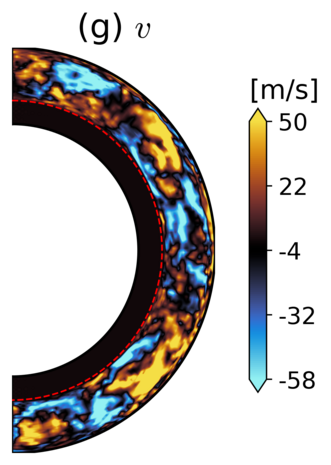}
    \includegraphics[width=0.17\textwidth]{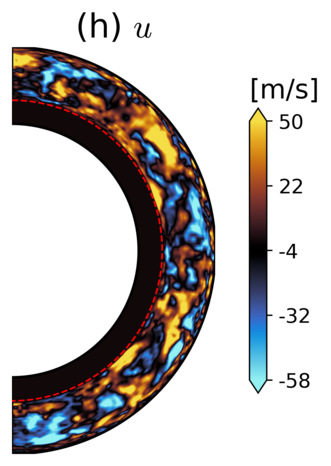}\\
    \includegraphics[width=0.27\textwidth]{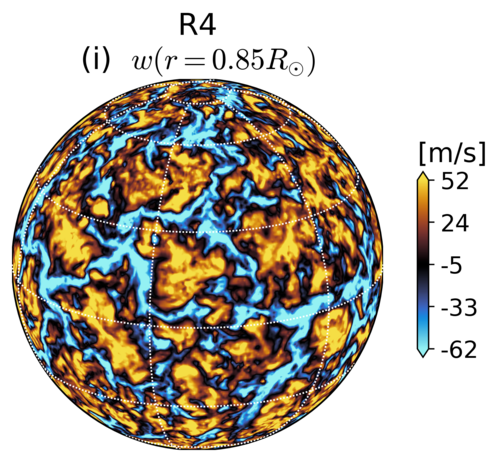} 
    \includegraphics[width=0.17\textwidth]{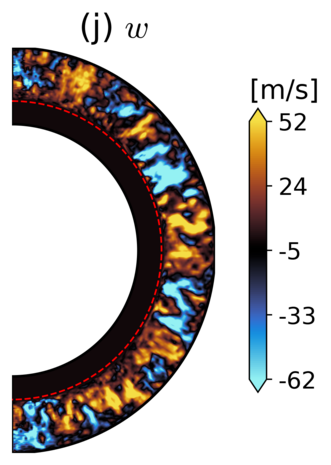}
    \includegraphics[width=0.17\textwidth]{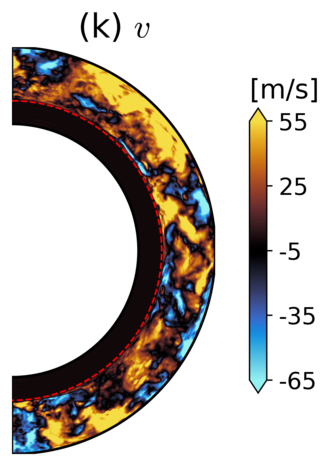}
    \includegraphics[width=0.17\textwidth]{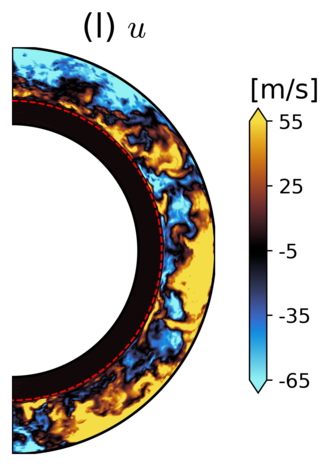}\\
    \includegraphics[width=0.27\textwidth]{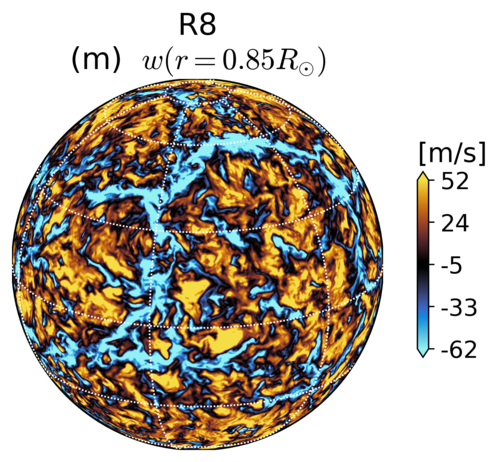} 
    \includegraphics[width=0.17\textwidth]{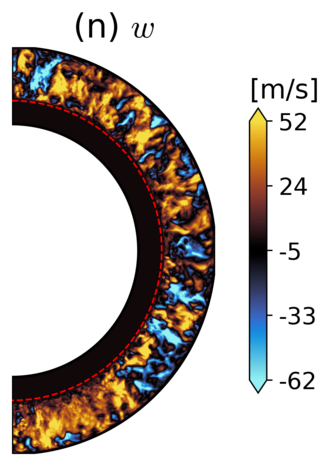}
    \includegraphics[width=0.17\textwidth]{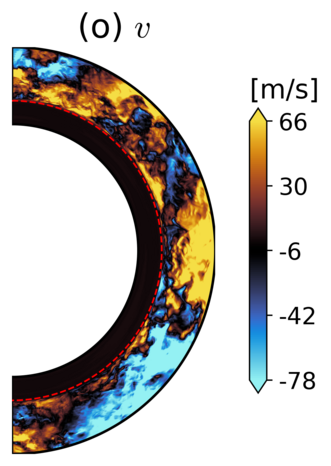}
    \includegraphics[width=0.17\textwidth]{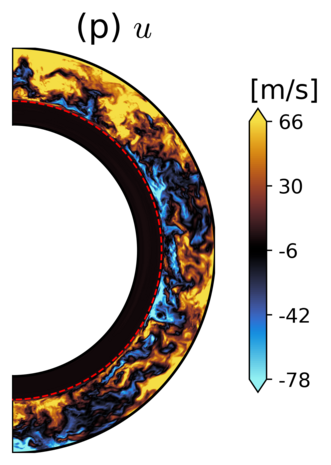}\\
    \caption{Instantaneous snapshots of the velocity components for 
      simulations R1 to R8, from top to bottom.  The radial velocity, $w$ is 
      presented in the orthographic projection  (panels a, e, i, and m),
	    and in the meridional plane (b, f, j, and n). The latitudinal, $v$ 
      (panels c, g, k, and o) and the longitudinal, $u$ (panels d, h, l, and p)
	    components are presented in the meridional plane. The thin dashed red line 
      shows the transition between the stable and unstable layers.}
    \label{fig:wvtsnap}%
    \end{center}
\end{figure*}

\begin{figure}%
    \begin{center}
    \includegraphics[width=1\columnwidth]{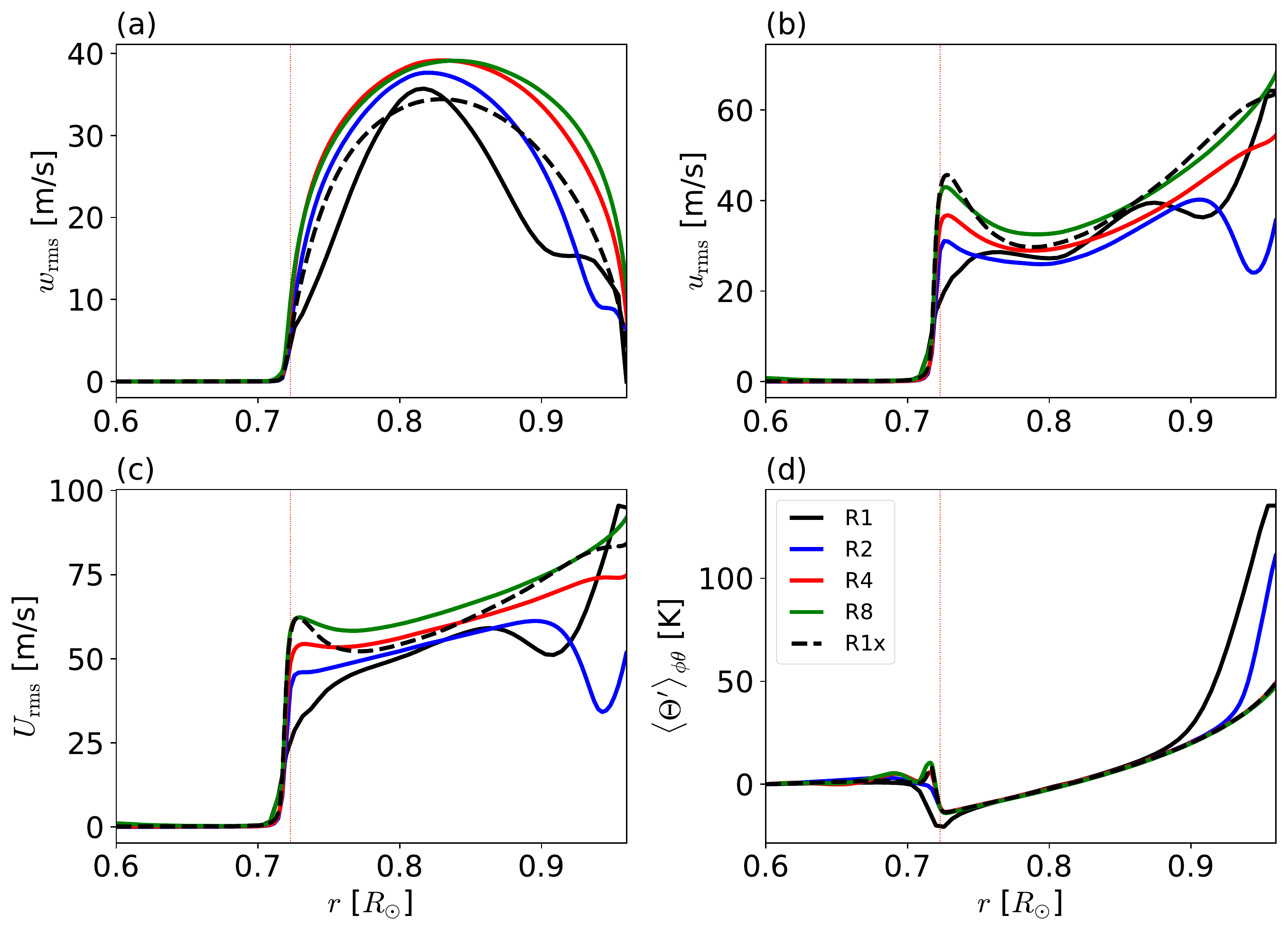} 
	    \caption{Radial profiles of $w_{\rm rms}$ (a),  
	    $u_{\rm rms}$ (b), the total $U_{\rm rms}$ (c); and the perturbations of 
      potential temperature (d). The profile of the latitudinal component 
	    $v_{\rm rms}$ is similar to $u_{\rm rms}$. The averaging was performed in 
      the horizontal directions and time. The thin vertical lines show the bottom of
      the unstable layer.}
    \label{fig:prof_r}%
    \end{center}
\end{figure}

We first explore the effects of numerical resolution in the low-order
moments and the spectral properties of non-rotating simulations.  
Figure~\ref{fig:wvtsnap} shows snapshots of the radial, latitudinal,
and longitudinal 
velocities, $w$, $v$, and $u$, for simulations R1-R8 (from top to bottom).
The graphs show the level of detail reached by different resolutions. 
The amplitude of the
velocity components increases as evidenced by brighter colors from the top to
the bottom panels. It is also evident that in models R1 and R2 the grid size
in regions with small density scale height, i.e., close to the model's top, is
insufficient to resolve the scales of the sub-surface motions. 
This is observed in the RMS profiles of the radial and longitudinal
components of the velocity field, 
$w_{\rm rms} = \sqrt{\brac{w^2}_{\phi\theta}}$ and $u_{\rm rms} = 
\sqrt{\brac{u^2}_{\phi\theta}}$,
as well as in the total RMS velocity, 
$U_{\rm rms} = \sqrt{\brac{u^2 + v^2 + w^2}_{\phi\theta}}$. Throughout the paper, 
angular brackets $\brac{}_{\phi\theta}$  correspond to averages over the horizontal 
directions and time, $\brac{}$ to volume and time averages over the convective shell, 
and overline corresponds to average over longitude and time. 
In $w_{\rm rms}$ the lack of resolution appears as flat profile
for $r > 0.9 \Rs$ for simulation R1 and for $r > 0.93 \Rs$ for R2, 
see Fig.~\ref{fig:prof_r}(a).  
In $u_{\rm rms}$ and  $U_{\rm rms}$, it appears as
a local minimum that may be observed in the profiles of the 
same simulations close to the upper boundary,
Fig.~\ref{fig:prof_r}(b and c).
This undesirable property was reported in the 2D Cartesian cases \citep{nogueira+22}. 
The black dashed line presented in the figure corresponds to simulation R1x,
it has the same horizontal resolution as R1, but with a fourfold resolution in the
radial direction ($N_{\phi}=128$,$N_{\theta}=64$,$N_r=256$). While there are small 
departures from the 
cases with higher resolution, especially in the radial  velocity, 
it is remarkable the similarity between case R1x and high resolution cases 
R4 and R8. 

The spherical simulations show radial motions which have a maximum roughly
at the center of the convection zone ($r\sim0.85\Rs$). The amplitude of 
$w_{\rm rms}$ slightly increases from the low to the high resolution cases,
with a rather small difference between case R4 and R8. 
The longitudinal velocity, $u_{\rm rms}$, shows larger amplitudes at the bottom 
of the convection zone resulting from the encounter of fast downward plumes 
with the rigid stable layer. As a consequence, the profiles 
of $u_{\rm rms}$ as a function of radius have a minimum in the bulk of the 
convection zone at about $0.8\Rs$. This feature does not appear in simulations 
that consider only the
convection zone, where a stress-free boundary condition is imposed at their 
base \citep[e.g.,][]{fan+14,hotta18}. It is important because it creates
a steeper transition between convective and radiative layers. 

\begin{figure}%
    \begin{center}
    \includegraphics[width=0.8\columnwidth]{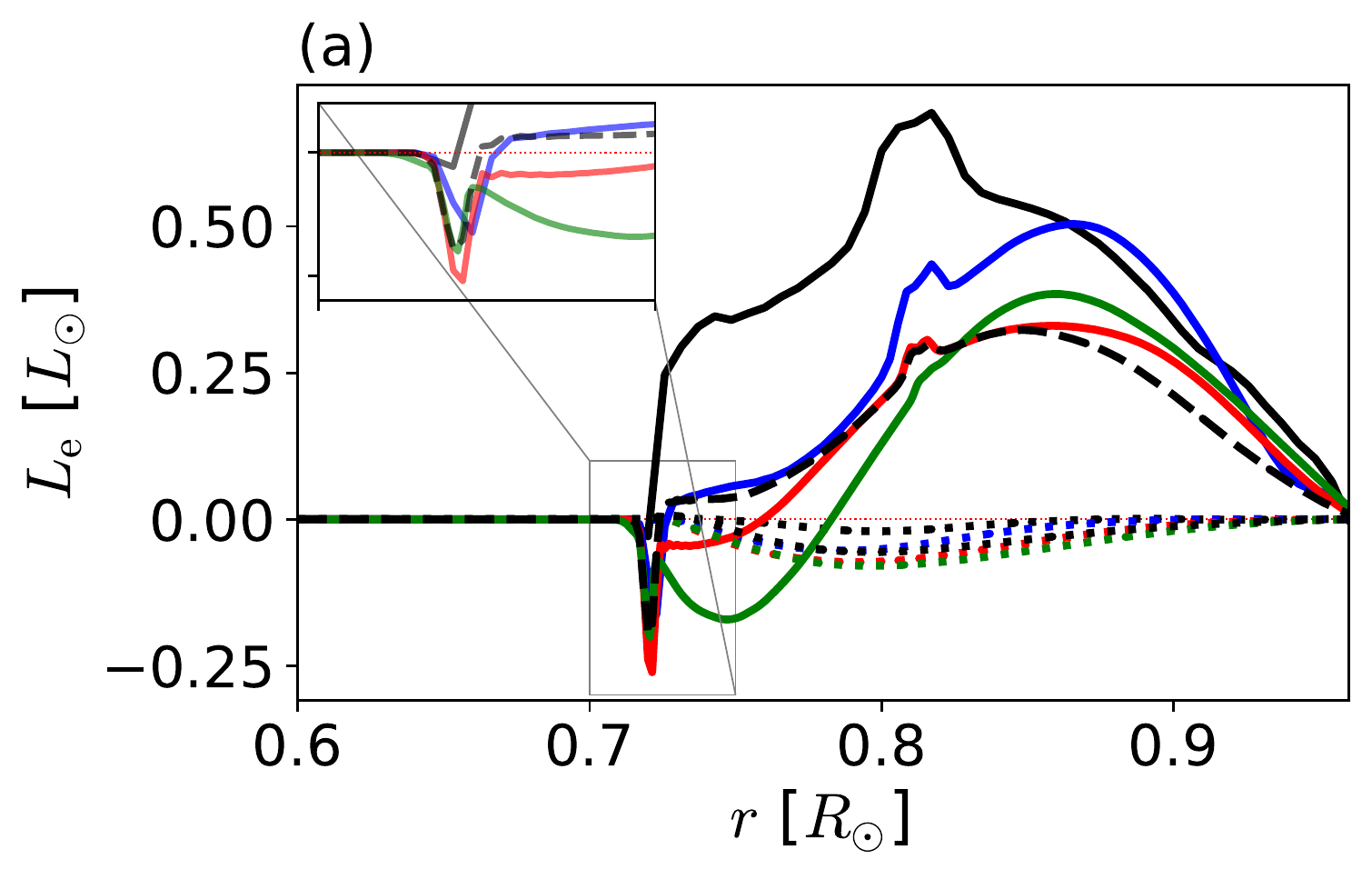} 
	    \caption{Luminosity carried by the enthalpy flux, 
	    $L_{\rm e} = 4 \pi r^2 R_g c_p \rho \brac{w \Theta^{\prime}}_{\phi\theta}$ 
	    (solid lines), and the kinetic energy flux, 
	    $L_{\rm k} = 4 \pi r^2 \rho \brac{w(u^2 + v^2 + w^2)}_{\phi\theta}$
	    (dotted lines) for the non-rotating simulations
      presented in Table~\ref{table:1}. The color coding of the lines is the 
      same as in Fig.~\ref{fig:prof_r}. The inset shows the overshooting region
      with negative values of $L_{\rm e}$. }
    \label{fig:cflux}%
    \end{center}
\end{figure}

The averaged values of the perturbations of potential temperature are presented
in Fig.~\ref{fig:prof_r}(d).  Despite simulations R1 and R2, which show higher
values of $\Theta^{\prime}$ at the upper levels, the profiles
for models R1x, R4 and R8 are rather similar. 
The profile of the luminosity, $L_{\rm e} = 4\pi r^2 F_{\rm e}$, carried by 
the enthalpy flux,  
$F_{\rm e} = R_g c_p \rho \brac{w \Theta^{\prime}}_{\phi\theta}$,  
for all non-rotating
cases, is shown with solid lines if Fig.~\ref{fig:cflux}(a). 
The profiles are normalized to 
the solar luminosity, $L_{\odot}$. Notice that $L_{\rm e}$ decreases with the 
increase  of resolution, and has a similar profile for simulations R1x and R4. 
The higher values reached in the low resolution simulations reflect the strong
perturbations of $\Theta$ in the upper layers of the domain in 
Fig.~\ref{fig:prof_r}(d).  The profile for case 
R8 is different perhaps due to insufficient statistics.  
The inset of the Fig.~\ref{fig:cflux}(a) shows the bottom of the convection 
zone where the 
enthalpy flux is negative. This thin layer, where the correlations between
$w$ (mostly negative due to fast downward plumes)  and $\Theta^{\prime}$ 
(positive) are negative, is associated with overshooting.  It is roughly zero 
for case R1 and increases with the grid size until reaching roughly similar values 
for the higher resolution simulations R4 and R8. The luminosity 
carried by the kinetic energy flux,  $L_{\rm k} = 4\pi r^2 F_{\rm k}$, with
$F_{\rm k}=\rho \brac{w(u^2 + v^2 + w^2)}_{\phi\theta}$ 
is presented in the same panel with dotted lines. 
The flux is negative for all cases with minimum values reached by the high 
resolution simulations. 

\subsubsection{Spectral analysis}
In this section we present the spectral properties of the non-rotating simulations.
We are interested in the distribution of energy among convective scales for the
different components of the flow. The library SHTns \citep{shtns}
\footnote{See also
\url{https://www2.atmos.umd.edu/~dkleist/docs/shtns/doc/html/index.html}}, 
is used to compute  the spectral representation  of the vector velocity field as,
\begin{eqnarray}
	{\bf u} (\theta,\phi) = \sum_{\ell} \sum_{m=-\ell}^{\ell} 
	& & q_{\ell}^m {\bf Q}_{\ell}^m(\theta,\phi) \\\nonumber
	&+& s_{\ell}^m {\bf S}_{\ell}^m(\theta,\phi) \\\nonumber
	&+& t_{\ell}^m {\bf T}_{\ell}^m(\theta,\phi),
\label{eq:vfieldsh}
\end{eqnarray}
where, 
\begin{eqnarray}
{\bf Q}_{\ell}^m &=& Y_{\ell}^m {\bf \hat{e}}_r \\
{\bf S}_{\ell}^m &=& r {\bf \nabla} Y_{\ell}^m  \\
{\bf T}_{\ell}^m &=& - {\bf r} \times  {\bf \nabla} Y_{\ell}^m \;,
\end{eqnarray}
and $Y_{\ell}^m$ are spherical
harmonics of degree $\ell$ and order $m$, with $-\ell \le m \le \ell$.
The coefficients $q_{\ell}^m$,  $s_{\ell}^m$ and $t_{\ell}^m$ 
are the radial and transverse components of the velocity vector.
Under this decomposition, the spherical harmonic representation of the 
total kinetic energy as a function of the harmonic degree $\ell$ is given by 
\begin{equation}
	\tilde{E}(\ell) = \sum_{m=-\ell}^{\ell} |q_{\ell}^m|^2 +
	\ell(\ell +1) \left(|s_{\ell}^m|^2 + |t_{\ell}^m|^2 \right)  \;.
\label{eq:kspec}
\end{equation}

\begin{figure*}%
    \begin{center}
    \includegraphics[width=\textwidth]{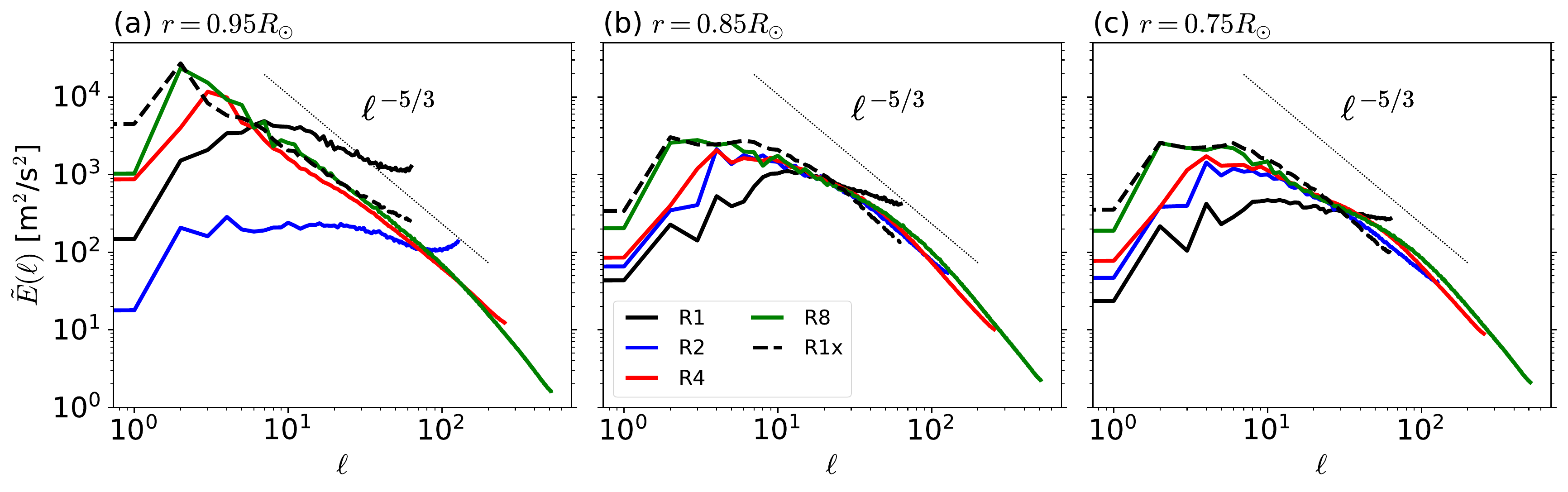} 
    \caption{Turbulent kinetic power spectra for simulations with 
	    different resolutions R1-R8 (see color correspondence in the legend), at
	    three different depths, from left to right. The black dashed line
	    corresponds to a simulation with the same horizontal resolution than 
      R1 but 256 grid points in the radial direction. The dotted line shows
	    the $\ell^{-5/3}$ Kolmogorov scaling.}%
    \label{fig:spectra}%
    \end{center}
\end{figure*}

The kinetic energy spectra for all non-rotating simulations (see legend in the figure)
are presented in Fig.~\ref{fig:spectra} for three different depths, 
(a) $r=0.95\Rs$, (b) $r=0.85\Rs$, and (c) $r=0.75\Rs$.  As it 
could have been anticipated from  the results in physical space, as a consequence 
of poorly resolved motions the simulations R1 and R2 have less kinetic energy 
at the upper layers, and show an excess of energy at large wave numbers.  
However, simulations R4 and R8 have a similar spectra with maximum amplitude at 
$\ell \sim 3$.  In the middle and bottom of the convection zone the energy at the
largest scales, $\ell < 4$, increases with the resolution. This indicates
that the effective viscosity is decreasing and its action shifting towards the 
smaller scales.  For $\ell \gtrsim 4$,  the kinetic spectra 
of simulations R2-R8 are similar. With the increase of the resolution the
spectra extend over more scales, reaching about two decades of $\ell$ for R8.

As for the energy cascade from the most energetic scales towards the
dissipative scales, the simulations show a scaling slower than 
the Kolmogorov law, $k^{-5/3}$ \citep{k41}, especially in the middle and bottom of the
convection zone. This result is expected given the buoyant force in 
an atmosphere with significant density stratification. Qualitative inspection
shows that the Kolmogorov rule seems to exist in small regions 
of the inertial scale. 

A surprising aspect of the figure is the spectra from simulation R1x presented 
with black dashed lines. Although this case has a coarse horizontal resolution, 
equivalent to R1, and high resolution, equivalent to case R4, only in radius, 
its kinetic power spectra are compatible with the highest resolution cases. 
The comparison shows that simulation R1x has an 
excess of energy in the largest scales.
However, the energies at the inertial range as well as the turbulent
scaling have good agreement with R4 and R8.

To assess the relevance of the present results, we compare them 
with solar supergranulation motions, 
which are weakly influenced by rotation. Although their nature is
still controversial, recent works point toward a buoyant convective origin
\citep{cossette+16,rincon+18}. Recently, \citet{rincon+17} developed analytical 
scaling laws for the spectral behavior of a flow in a stratified atmosphere 
in the presence of buoyancy, i.e., anisotropic turbulence.  
They successfully compared these laws with the spectra of supergranulation 
reconstructed from Doppler and photometric measurements of the HMI instrument on
the SDO satellite \citep{scherrer+12}, and decomposed in the spherical harmonic 
components. This supports the
hypothesis that supergranulation is a particular scale  of buoyantly driven 
convection. It is possible to  evaluate whether
the convectively driven motions developed in the simulations above 
compare with their results.

\begin{figure}%
    \begin{center}
    \includegraphics[width=\columnwidth]{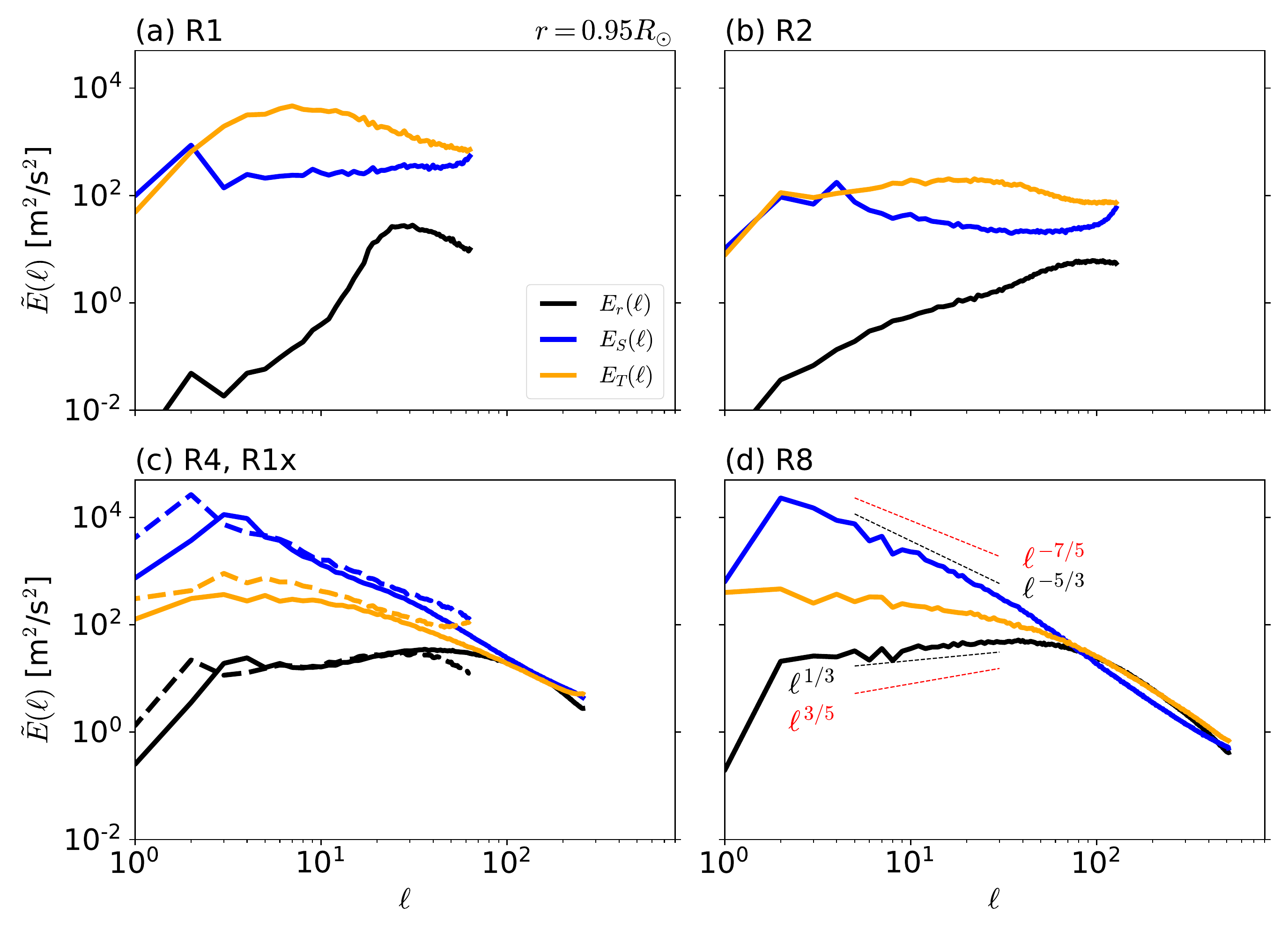} 
	    \caption{Kinetic power spectra for the spheroidal (blue lines), 
	    toroidal (yellow) and radial (black) components of the velocity
	    field at $r=0.95\Rs$ for simulations (a) R1 to (d) R8. 
	    The dashed lines in panel (c) correspond 
      to simulation R1x.  The results of panels (c) and (d) indicate strongly 
      anisotropic convection with divergent flows more energetic than the radial 
      flows. The black and red dashed lines in panel (d) show the 
      \cite[][black dashed lines]{k41} and \citet[][red lines]{rincon+17} 
      predictions for the spectra of turbulent motions.}%
    \label{fig:RSTspectra}%
    \end{center}
\end{figure}

Fig.~\ref{fig:RSTspectra} shows the kinetic energy spectra decomposed into 
the radial (black lines), spheroidal (blue) and toroidal (yellow)  
components at $r=0.95\Rs$ for simulations R1 (panel a)
to R8 (d), and R1x (presented in panel c). These components are obtained by 
separating the terms in the RHS of Eq.~(\ref{eq:kspec}),
\begin{eqnarray}
\label{eq:RSTspec}
  \tilde{E}_Q(\ell) &=& \sum_{m=-\ell}^{\ell} |q_{\ell}^m|^2 \;, \\\nonumber
  \tilde{E}_S(\ell) &=& \sum_{m=-\ell}^{\ell} \ell(\ell +1) |s_{\ell}^m|^2\;, \\\nonumber
  \tilde{E}_T(\ell) &=& \sum_{m=-\ell}^{\ell} \ell(\ell +1) |t_{\ell}^m|^2  \;.
\end{eqnarray}
Under this decomposition, $\tilde{E}_S$ and $\tilde{E}_T$ are measurements of the
flow divergence and vorticity in the ($\phi$,$\theta$) plane, respectively. 
In this sense, toroidal does not correspond to the longitudinal component 
of the flow, as usually assumed in mean-field theory.
In this representation, there is a striking difference of simulations 
R1 and R2  from all the others.
These simulations have more energy in the toroidal motions than in th radial 
and spheroidal motions. The distribution of energy changes in 
simulation R2, but a convergent pattern seems to appear for the cases R4 and R8. 
The results clearly show anisotropic turbulence, with the spheroidal, divergent,
part of the motions having $\sim10^3$ times more energy than the radial flows. In 
panel (c), simulations R4 (continuous lines) and R1x (dashed lines) are compared.
Both cases show similar spectral properties. In panel (d) the
red dotted lines correspond to the scaling laws for spheroidal
and radial motions derived by \citet{rincon+17}. The black dotted lines correspond to 
their Kolmogorov equivalents.   The scaling laws followed by the motions in our
simulations R4, R8 and R1x, certainly of buoyant origin,  compare well with those 
of \citet{rincon+17}. However, the Kolmogorov laws are also plausible, and it 
is hard to distinguish what law is followed and in what range of the spectra.

Caution is needed in this  comparison because the simulations correspond to 
a thick shell and develop scales 
considerably larger than supergranulation (supergranulation has its maximum energy at 
$\ell \sim 120$, whereas the simulations results show peaks at $\ell\sim3$). 
Nevertheless, the energy distribution among the different components of the 
velocity, the morphology of the spectral curves, with the spheroidal motion 
peaking at a large scale, the radial kinetic energy having a kink at the same 
scale but peaking at a smaller one, and the compatible turbulent scaling laws, 
demonstrate that simulations R1x, R4 and R8 capture well the properties of 
convective motions unconstrained by rotation. 

It is also worth comparing the findings described above with those
of other in similar simulations.
The spherical high resolution simulation, case SD,  
presented in  \citet{hotta+19} has a radial profile of the RMS velocity that is
qualitatively similar to the profiles of $U_{\rm rms}$ presented
in Fig.~\ref{fig:prof_r}(c). Because in their simulations convection
carries almost the entire solar luminosity, the amplitude of the velocity
is larger. The turbulent kinetic spectrum of this case has
about two orders of magnitude more energy in the horizontal than in the
radial motions \citep[see also][]{hotta+14}. 
In the cases presented here, the anisotropy is larger, as can be seen
in Fig.~\ref{fig:RSTspectra}(c and d).
In their spectrum for the horizontal velocity, the maximum
is at $\ell \sim 5$ or $6$, in fair agreement with our findings.  

In the study performed by \citet{featherstone+16a}, including low-and 
high-resolution simulations, the situation is different.
Their sets of simulations with density scale heights $N_{\rho}=3$ and
$4$ are compatible with the experiments presented here.  
Unlike what is observed in Fig.~\ref{fig:prof_r}(a), in their profiles
of the RMS radial velocity component the maximum shifts 
towards the top boundary of the model with the increase of the resolution 
(also larger Rayleigh number). In their
kinetic spectra, the largest scale $\ell=1$, has the maximum 
power. In their low-resolution simulations, the power at low
$\ell's$ is larger and decreases for higher resolution, where the
newly resolved motions acquire considerable energy. Thus, their
inertial range becomes flat. This difference is intriguing. It 
might arise from the difference in the energy equation between
this work and \citet{featherstone+16a}. However, both 
\citet{featherstone+16a}
and \citet{hotta+19} consider somewhat similar energy equation, 
including a static background state and radiative diffusivity.
On the other hand, \citet{featherstone+16a} consider explicit viscous
dissipation, whereas in this work and in \citet{hotta+19} it comes
from the numerical scheme.  

\subsection{Rotating solar convection}
\label{sec:rot}
\begin{figure*}%
    \begin{center}
    \includegraphics[width=0.27\textwidth]{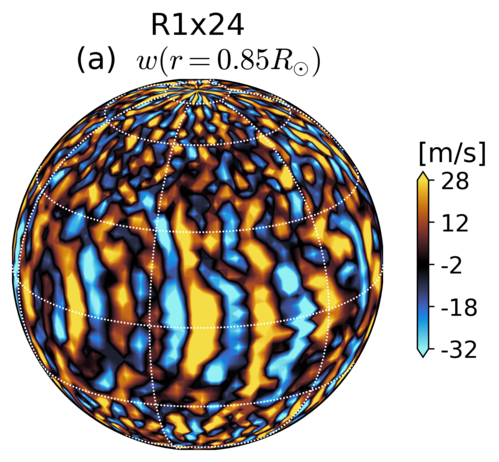} 
    \includegraphics[width=0.17\textwidth]{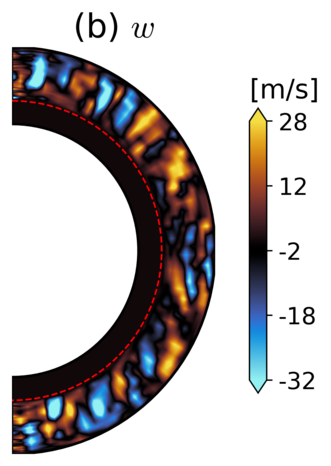}
    \includegraphics[width=0.17\textwidth]{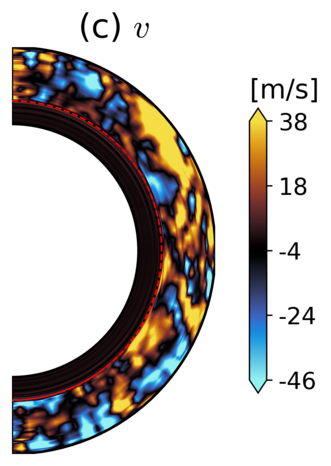} 
    \includegraphics[width=0.17\textwidth]{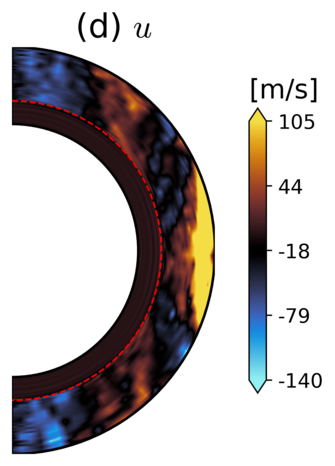}\\
    \includegraphics[width=0.27\textwidth]{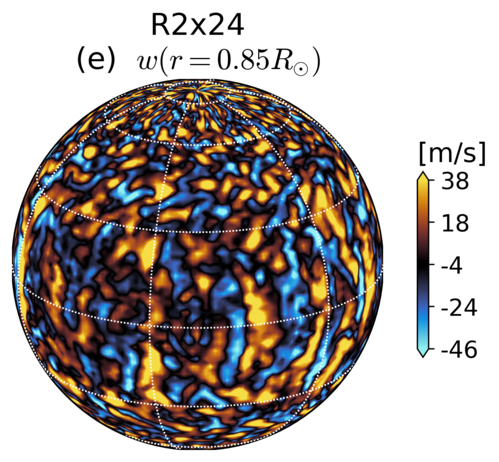} 
    \includegraphics[width=0.17\textwidth]{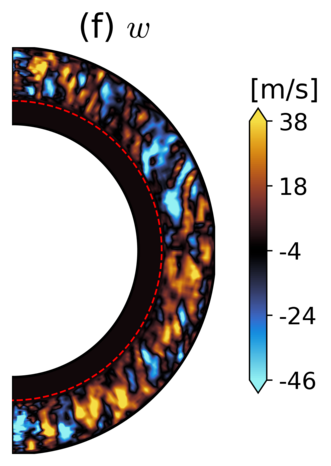}
    \includegraphics[width=0.17\textwidth]{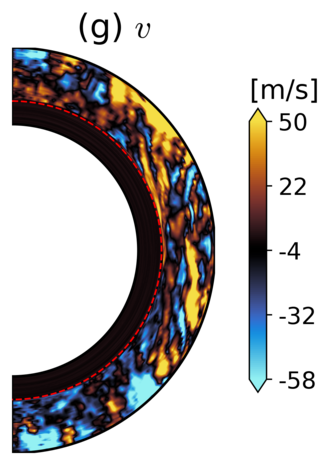} 
    \includegraphics[width=0.17\textwidth]{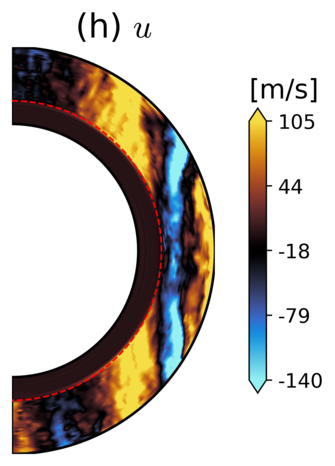}\\
    \includegraphics[width=0.27\textwidth]{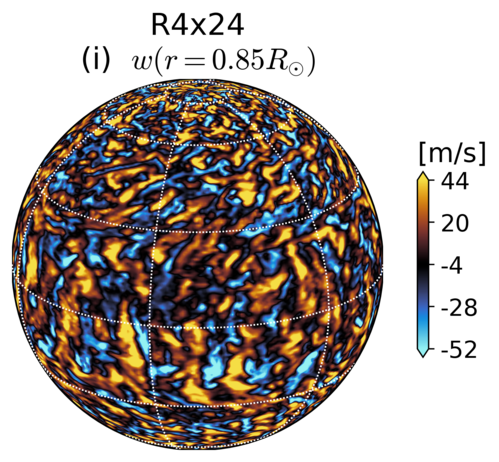} 
    \includegraphics[width=0.17\textwidth]{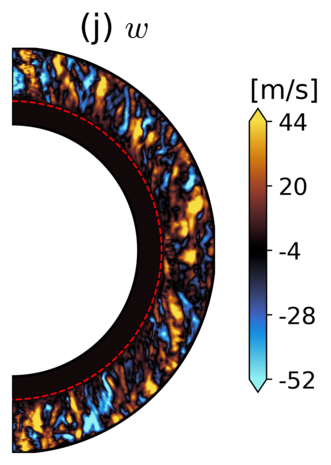}
    \includegraphics[width=0.17\textwidth]{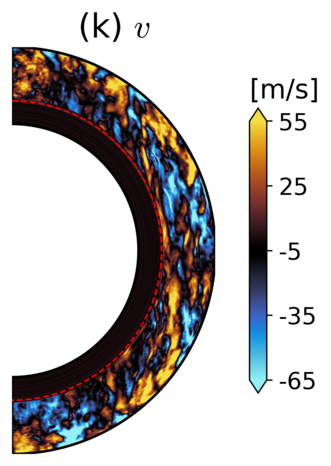} 
    \includegraphics[width=0.17\textwidth]{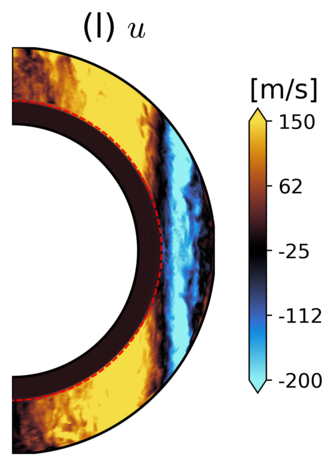}\\
      \caption{Same as Fig.~\ref{fig:wvtsnap} for simulations R1x24, R2x24 and R4x24, 
      from top to bottom.}
    \label{fig:snap_rot}%
    \end{center}
\end{figure*}

Here we present simulations that 
include the Coriolis force in the momentum equation, Eq.~(\ref{eq_momentum_nc})
(see Table~\ref{table:1}).
In numerical experiments not presented here, we noticed that the convergence to a 
steady state solution
occurs faster if the initial conditions are random perturbations rather than the 
non-rotating relaxed state. 
Figure~\ref{fig:snap_rot} depicts the morphological characteristics of the
instantaneous radial flow in the orthographic projection at $r=0.85\Rs$ (see a,e and i 
panels for simulations R1x24, R2x24 and R4x24, respectively), and in the meridional 
plane (b,f,j); the latitudinal velocity in the meridional plane (c,g,k), and the 
longitudinal velocity (d,h,l). 

The orthographic projection shows the elongated
structures at equatorial latitudes also known as ``banana cells", characteristic
of rotationally constrained motions taking the form of columns \citep{busse86}. 
For the cases R2x24 and R4x24 regions of strongly stretched structures are 
observed at intermediate latitudes. These are a consequence of the resulting 
large-scale shear.  In the meridional plane,  
the convective motions at equatorial latitudes are not radial as in the non-rotating
cases presented in Fig.~\ref{fig:wvtsnap}, but elongated convective columns aligned
with the rotation axis. At higher latitudes, there are multiple convective cells,
exceeding the number in the non-rotating cases, with a certain tilt regarding the 
rotation axis.  The latitudinal velocity (panels c, g and k)  shows  
flow parcels that are aligned to the rotation axis. The number of these 
structures seem to increases with the resolution.  Finally, the instantaneous
snapshot of the longitudinal velocity evinces the sustained large-scale longitudinal
flow resulting from each case.  The low-resolution case shows an accelerated equator
and deceleration towards high latitudes (panel d).  The case R2x24 shows a fast  
rotating equator, a column of retrograde flow above the tangent cylinder, and a column of 
accelerated motions inside the tangent cylinder (h).  In the case R4x24, the 
velocity at the equator diminishes, whereas the higher latitude acceleration
increases and advances towards the poles (l).  The sustainment of this longitudinal 
mean-flow, and its coupled meridional flow complement are discussed in 
\S\ref{sec:amt}. 

\begin{figure}%
    \begin{center}
    \includegraphics[width=\columnwidth]{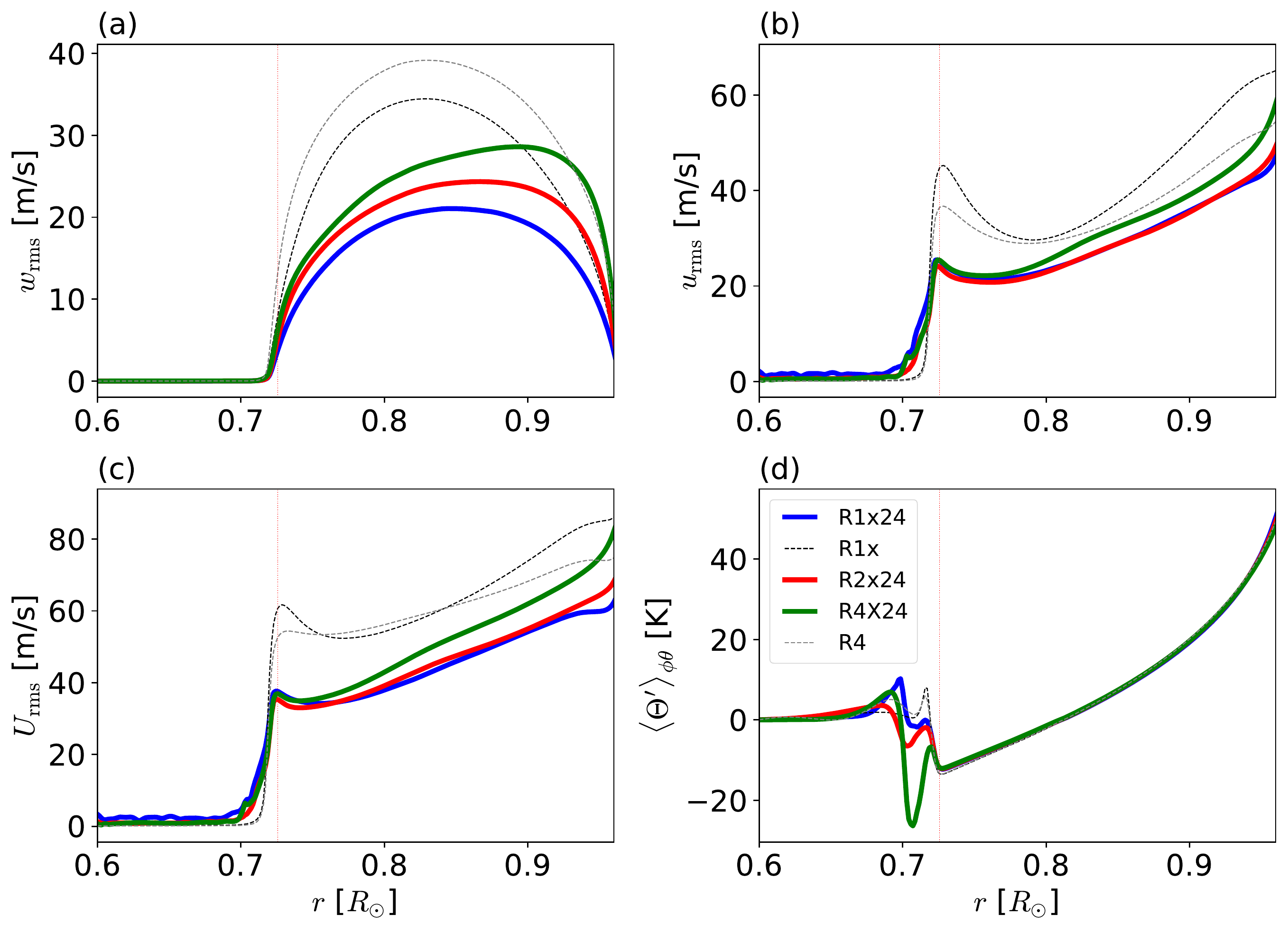} 
    \caption{Same as Fig.~\ref{fig:prof_r} but for the rotating simulations with
      different resolutions, R1x24 (blue), R2x24 (red) and R4x24 (green). For
      comparison, the profiles corresponding to the non-rotating cases R1x and 
      R4 are presented with black and gray dotted lines, respectively. The thin
      red vertical lines show the bottom of the convection zone.}
    \label{fig:prof_rot_r}%
    \end{center}
\end{figure}

Figure~\ref{fig:prof_rot_r} compares the same averaged quantities presented in 
Fig.~\ref{fig:prof_r} between the rotating cases, R1x24, R2x24 and R4x24 (blue,
red and green lines, respectively). For comparison, the results for the non-rotating
models R1x and R4 are presented in black and gray dotted lines, respectively. 
It is expected that the amplitude of the velocity field components is quenched
by the Coriolis force. This change is evident by comparing the continuous color 
lines with the dotted dark lines.  For simulations R1x24 and R4x24 the mean 
radial velocity, $w_{\rm rms}$ is 
$\sim 65 \%$ and $\sim 70 \%$ of the values for the cases R1x and R4,
respectively (panel a). Most importantly for this paper is the observed
trend of this quantity to acquire larger values with the increase of
numerical resolution.   Note also that  the maximum of $w_{\rm rms}$ 
moves towards the upper boundary.  The longitudinal component, 
$u_{\rm rms}$ (b), also decreases compared to the
models without rotation, but the change is not as significant as for 
$w_{\rm rms}$.  It is noteworthy that the RMS horizontal motions slightly 
penetrate into the stable layer at $r\sim0.7\Rs$ (see vertical red line), 
and even at deeper layers there is a non-zero velocity. This non-vanishing
velocity 
suggests that there is a weak level of turbulence in the stable layer.  
The perturbations of $\Theta$ (panel d) 
do not show significant changes in the convection zone with respect to the 
non-rotating cases.  Also, there are no notable differences between simulations 
with different  resolutions. However, at the bottom of the CZ, there is a 
valley of negative 
perturbations that is deeper for higher resolution.
A positive peak can be identified slightly below the CZ. Its location is roughly 
the same for all rotating cases. 
As seen below, these maxima appear as a consequence of the large-scale
shear that establishes at the tachocline and their associated thermal
wind balance. 

\begin{figure}%
    \begin{center}
    \includegraphics[width=0.8\columnwidth]{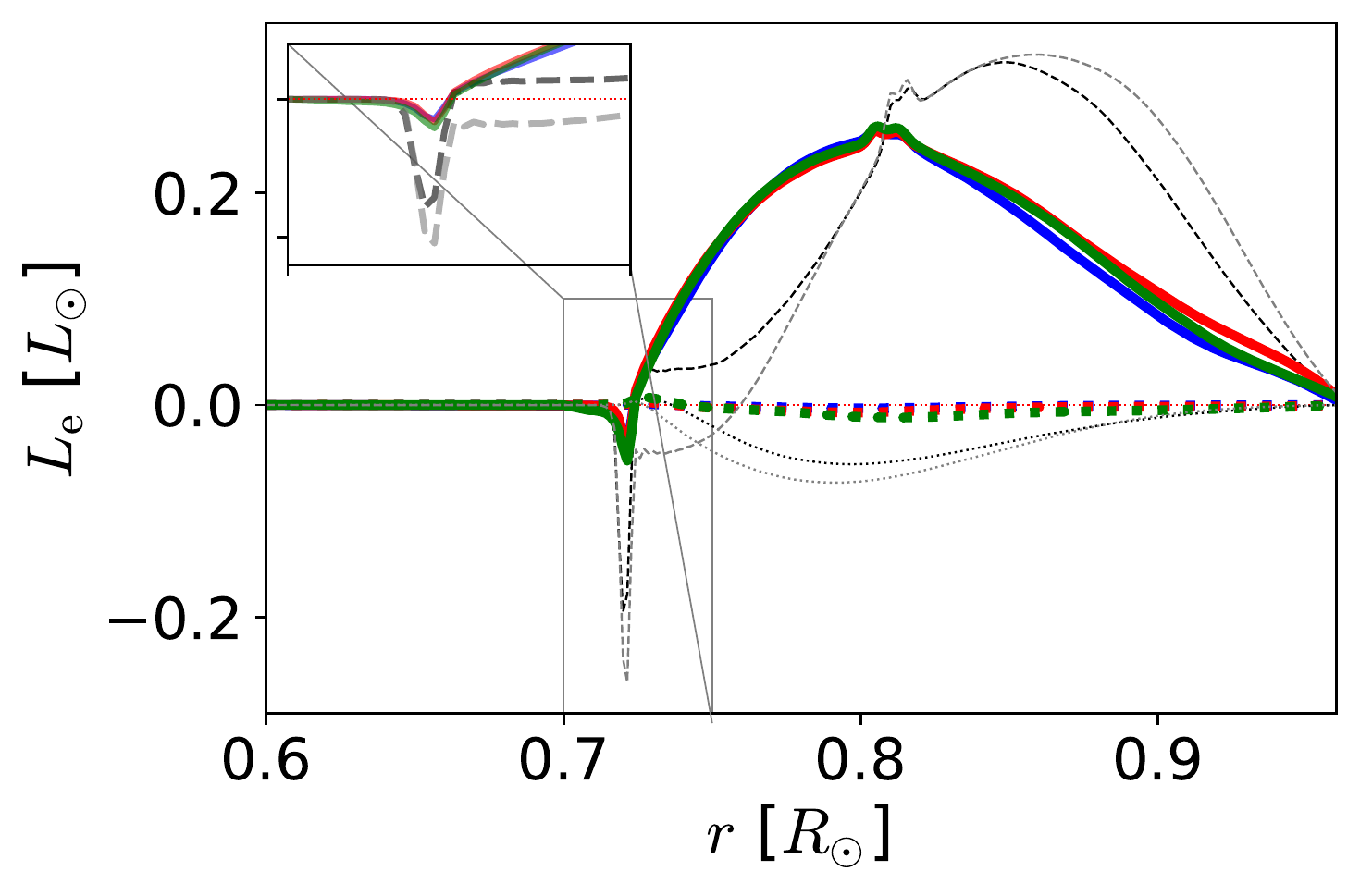} 
    \caption{Luminosity carried by the enthalpy flux in the rotating simulations
      R1x24, R2x24 and R4x24.  For comparison to the non-rotating cases, the dotted 
      black and gray lines 
      correspond to simulations R1x and R4, respectively. The inset focuses on the
      bottom of the convection zone and shows that overshooting decreases because
      of rotation. The dotted lines show the luminosity carried by the kinetic
      energy flux. }
    \label{fig:enth_flux_rot}%
    \end{center}
\end{figure}

The luminosity carried by the enthalpy flux resulting in the rotating 
simulations is presented in 
Fig.~\ref{fig:enth_flux_rot}.  It is interesting that, despite the differences
in $w_{\rm rms}$ between simulations R1x24 to R4x24, 
the average of the correlations $w_{\rm rms} \Theta^{\prime}$ 
remain roughly the same for all cases, carrying roughly $25\%$ of the solar
luminosity. Also, despite that rotation makes $w_{\rm rms}$ to 
peak closer to the top of the domain, the maxima of $L_{\rm e}$ is shifted 
downwards, with respect to the non-rotating cases (see for comparison, 
the black dotted lines in the figure, corresponding to simulations R1x and 
R4).  The figure also shows that rotation diminishes overshooting.
The bottom of the convection zone is focused in the inset in the figure. It
clearly shows a considerable difference between the non-rotating cases, presented
in dashed dark lines, and the rotating simulations. Finally, 
the dotted lines in the figures show the luminosity carried by the kinetic 
energy flux.  It is evident that it is much smaller than in the non rotating
cases. Therefore, the luminosity carried by convection, cf., the residual between
$L_{\rm e}$ and  $L_{\rm k}$, is mostly due to the enthalpy flux.\\

\subsubsection{Spectral analysis}

\begin{figure*}[htb]%
    \begin{center}
    \includegraphics[width=\textwidth]{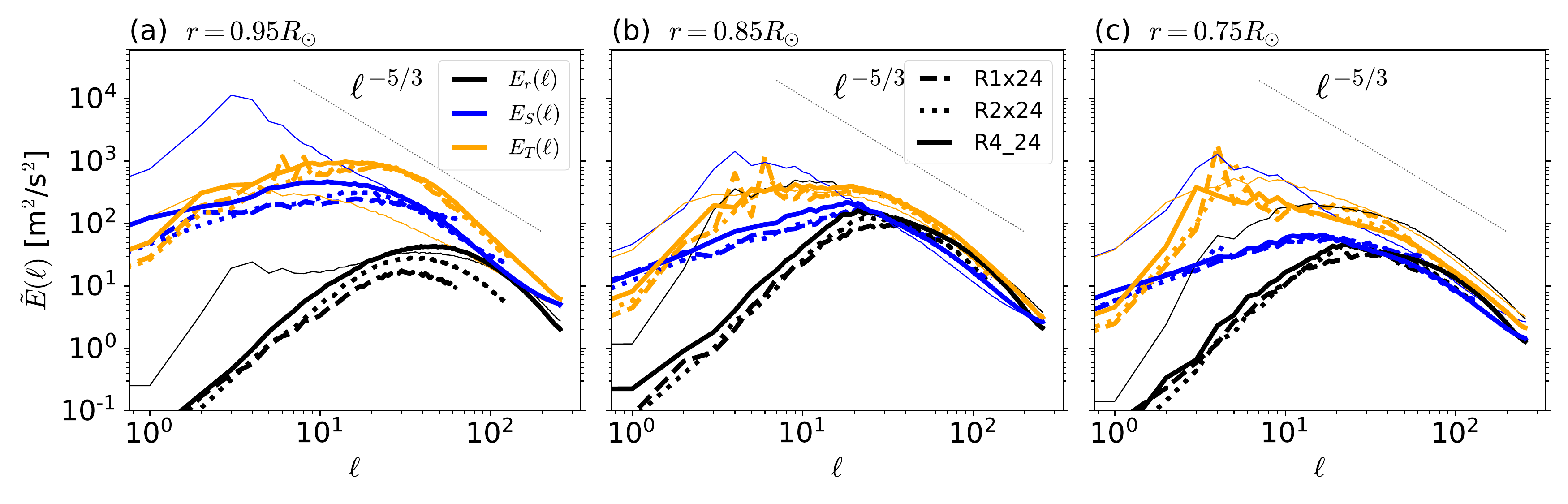}\\
	    \caption{(a) to (c) Same as Fig.~\ref{fig:RSTspectra} for 
	    simulations R1x24 (dashed line), R2x24 (dotted) and R1\_24 (solid). 
	    Note that the peak of the spectra of the rotating simulation 
	    is shifted towards the smaller scales, $\ell \sim 40$. 
	    Moreover, increasing resolution leads to larger power
	    in the radial component, with maximum being shifted to the 
	    smaller scales. For comparison, the thin solid lines  show
	    the spectra of case R4 at the corresponding depths.}%
    \label{fig:spectra_rot}%
    \end{center}
\end{figure*}

The turbulent kinetic energy spectra of the rotating simulations at different 
depths (a-c) are presented in
Fig.~\ref{fig:spectra_rot}. In this figure only the decomposition
in the $\tilde{E}_Q$ (black line), $\tilde{E}_S$ (blue) and  $\tilde{E}_T$ (yellow) 
components (Eq.\ref{eq:RSTspec}) is presented.  
The full kinetic spectra, (Eq.~\ref{eq:kspec}), follow 
the curves of the most energetic component, in this case, the toroidal kinetic 
energy. Note also that these spectra correspond to the turbulent velocity field, 
removing the axisymmetric component, $m=0$ from the total velocity vector, i.e., 
${\bf u}^{\prime} = {\bf u} - \mean{{\bf u}}$.  
The thick dashed, dotted and solid lines correspond to the
cases R1x24, R2x24 and R4x24, respectively. The spectra for the non-rotating
case R4 is presented in thin continuous lines for comparison. 
For rotating convection they also reveal anisotropic motions. This time,
however, the toroidal, component is the dominant one. The
radial velocity is the less energetic component. The anisotropy is
more pronounced near the top of the model domain, with an energy difference 
of about two orders of magnitude at the scales of the most energetic motions.  
In the middle and bottom of the CZ, the anisotropy diminishes and the energy 
difference is about a factor of two. 

As a consequence of the Coriolis force, the broad convective cells observed
for non-rotating convection, with maximum energy at the harmonic degree 
$\ell \sim 3$,  are broken into cells with scales peaking at $\ell \sim 30$ at 
the upper part of the domain and with most of the energy in $\tilde{E}_T$.  
At the middle and bottom there are peaks of energy at 
$\ell$'s between 6 and 8. 

It has been reported in both, HD \citep{featherstone+16b} and MHD simulations 
\citep{guerrero2019sets} that the spatial scale where the spectra  peak 
depends on the Rossby number. The results presented in this study show that the 
Rossby number also changes with the numerical resolution. However, a shift in the 
spatial scale of the energy peak is observed only for the radial component 
at the domain's surface (panel a).
$\tilde{E}_Q$ peaks at $\ell \sim 30$ for simulation R1x24, 
$\ell \sim 40$ for simulation R2x24, and $\ell \sim 50$ for case R4x24.
Concurrently, the energy increases by a factor of two to three between the low 
and the highest resolution cases. 
In the Sun, the radial spectral energy has a maximum at  $\ell \sim 400$ for
supergranular motions.  Yet, as previously mentioned, supergranules are mainly
divergent motions with most of the energy contained in the spheroidal energy 
component.  Interestingly,  a relevant change in 
the energy-containing scale of the horizontal velocity components with the
mesh size is not observed. It is worth remembering here that the longitudinal 
velocity which largely contributes to $\tilde{E}_T$  is the component used 
for measuring the observational spectra \citep{proxauf_phdT}. 
At large $\ell$'s the spectra decay with a scaling law 
faster than the Kolmogorov $\ell^{-5/3}$ rule at all depths (see the thin dotted lines).
It is a remarkable result that the spectra do not show dramatic changes
with resolution except the one described for $\tilde{E}_r$ at surface levels. 
Surprisingly, the large-scale patterns deriving from these turbulent
flows result in fully divergent outcomes, as will be presented below.

It is insightful to compare these findings with the 
results of the high resolution rotating simulation performed by \citet{miesch+08}. 
Their kinetic spectra show anisotropic motions at surface levels with
the maxima energy in the horizontal velocity components and at the harmonic
degrees $\ell = 20-30$. Their spectrum of the radial velocity component peaks 
at $\ell= 80$. Below the model top boundary their motions become more 
isotropic. These results are, in general, in good agreement with those presented in
Fig.~\ref{fig:spectra_rot}, i.e., with the increase of resolution the
energy of the radial velocity shifts to larger energy and smaller scales.
On the other hand, the spectra of the horizontal components seems to be
independent on the resolution.  Nonetheless, despite the agreement in
the spectral properties, the resulting mean-flows of simulation R4x24 
differ from their findings \citep[see Fig.~6 of][]{miesch+08}.

\begin{figure*}[htb]%
    \begin{center}
    \includegraphics[width=0.32\textwidth]{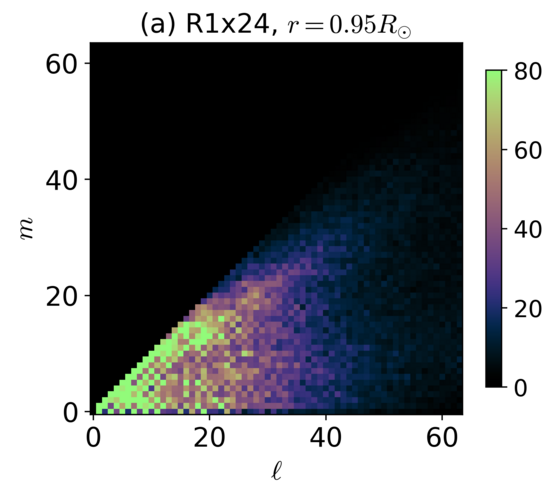} 
    \includegraphics[width=0.32\textwidth]{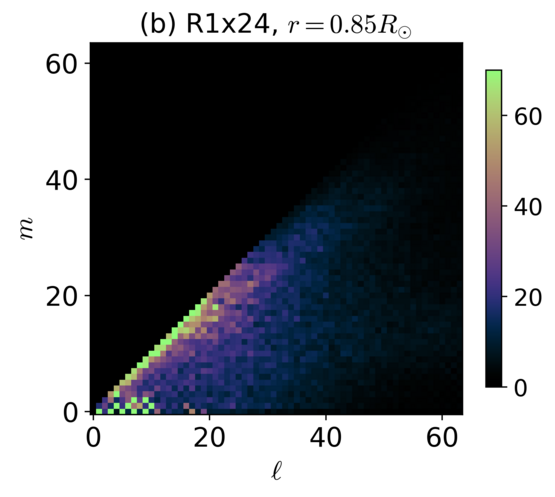} 
    \includegraphics[width=0.32\textwidth]{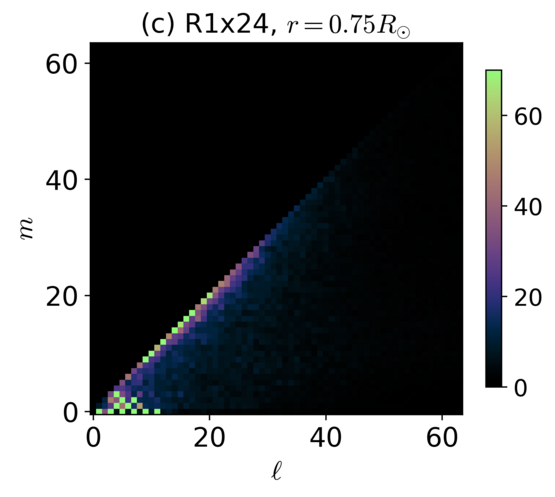} 
	    \caption{Two-dimensional kinetic energy spectra for the simulation
      R1x24 at the top, middle and bottom of the domain (a to c). The bottom left
      points observed in panels (b) and (c) correspond to inertial modes 
      driven by the Coriolis force.}%
    \label{fig:spectra_rot_2d}%
    \end{center}
\end{figure*}

From the graphs presented in Fig.~\ref{fig:spectra_rot} stand out the 
maximums appearing between 
$6 \lesssim \ell \lesssim 10$. They are more evident at the middle and 
the bottom of the domain, Fig.~\ref{fig:spectra_rot}(b and c).  As a matter 
of fact, these peaks contain the largest energy at the bottom of the CZ. 
To identify to what motions these peaks 
correspond, Fig.~\ref{fig:spectra_rot_2d} presents two-dimensional
spectra for the simulation R1x24. These spectra are computed using 
Eq.~(\ref{eq:kspec}) but not considering the sum over $m$ and averaging 
only over time. The panels (b) and (c) reveal that the 
high energy harmonics in this particular range correspond  to 
low-order longitudinal modes, $m=1$ to $4$. These are inertial modes, similar
to Rossby waves,  developing
below the convection zone.  It is observed that their maximal energy
is at depth $r=0.71\Rs$ and, quite remarkably, these peaks appear at the
same scales independently of the resolution. Because of their energy, 
which seems to propagate upwards to higher radial levels, these modes are
likely dynamically important in the rotating convective system. 
Assessing the properties of these waves and their relevance is left 
for an independent study  (Dias et al., in preparation).

The banana cells, clearly observed in Fig.~\ref{fig:snap_rot}(a, e, i) at 
$r=0.85\Rs$ 
have longitudinal wave numbers $16 <m <20$,
therefore in the $(m,l)$ plane they are close to the diagonal in
Fig.~\ref{fig:spectra_rot_2d}(b). At the top of the domain, 
the 2D spectrum shows large energies in the diagonal but also below
it. The analysis of solar motions by \citet{getling+22} 
presents similar 2D spectra.  At the deepest layers reached by their measurements,
$19$ Mm below the solar surface,  their spectrum shows higher energy 
levels below the diagonal with scales peaking between $10 \lesssim \ell \lesssim 40$.
The results presented in Fig.~\ref{fig:spectra_rot_2d}(a) resemble these
observations. 

\subsubsection{Mean-flows and angular momentum transport}
\label{sec:amt}

\begin{figure}%
    \begin{center}
    \includegraphics[width=\columnwidth]{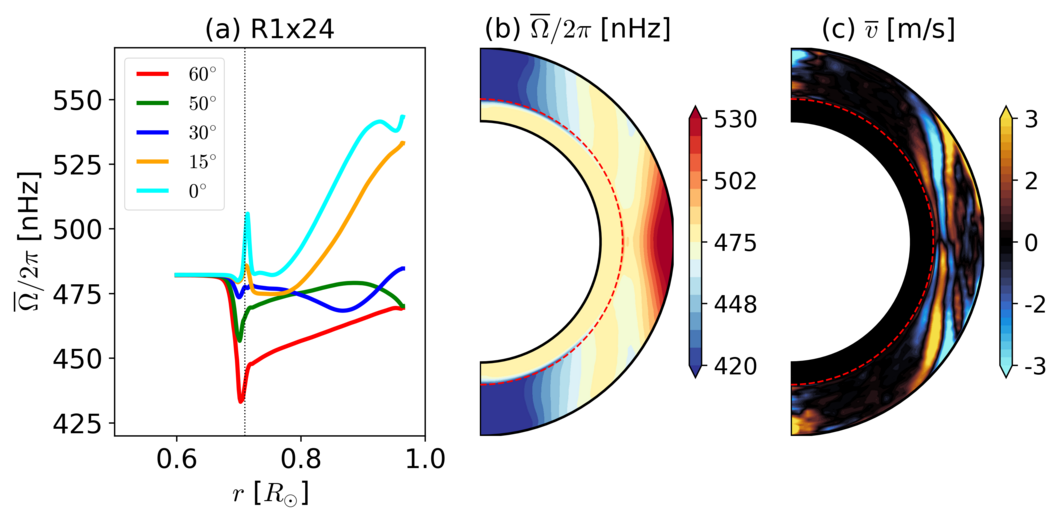} 
    \includegraphics[width=\columnwidth]{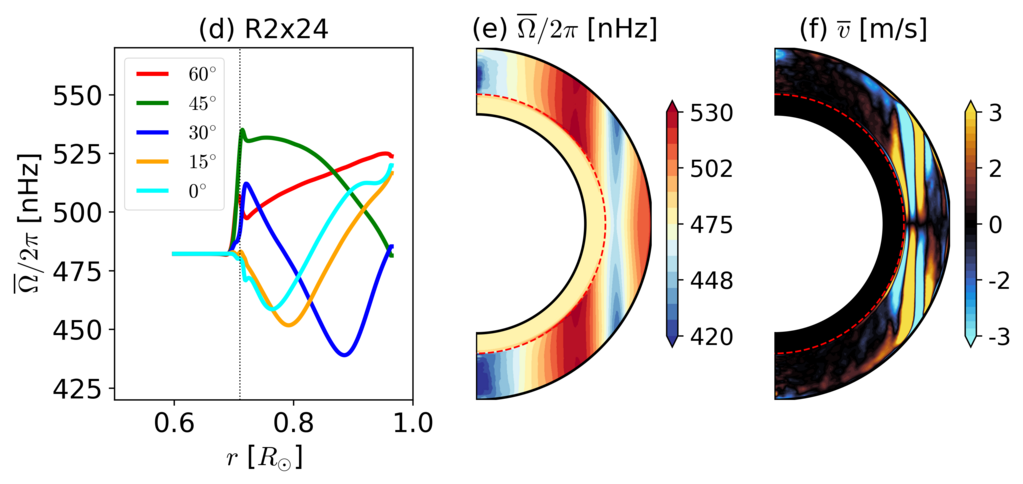}
    \includegraphics[width=\columnwidth]{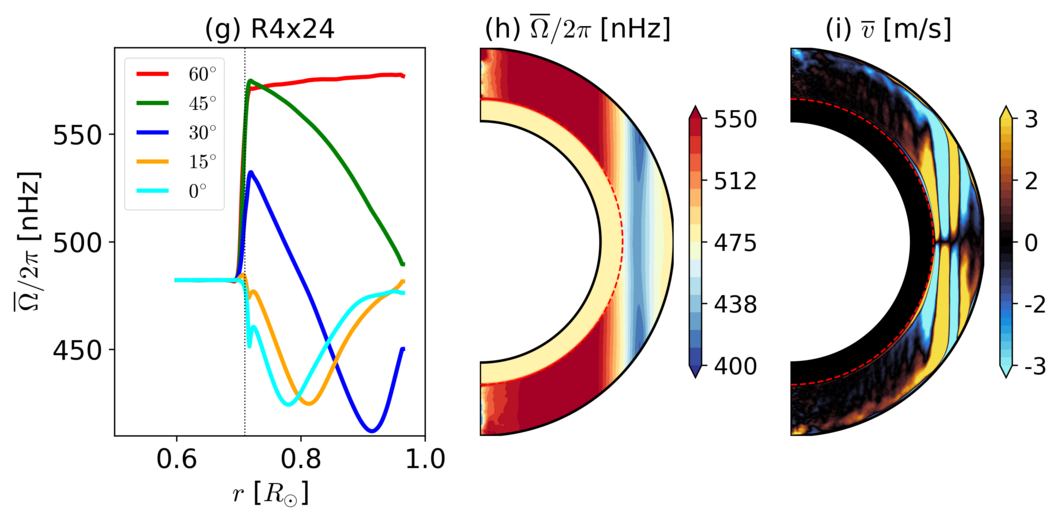}
      \caption{Left and middle panels: differential rotation as a function of
      radius for different latitudes, and in the meridional plane, respectively,
      for simulations  R1x24, R2x24, and R4x24, from top to bottom.
      Right panels: mean profile of the latitudinal velocity in the meridional plane. 
      In the northern hemisphere negative (positive) values correspond to poleward
      (equatorward) flows.}
    \label{fig:mean_flows}%
    \end{center}
\end{figure}

Turbulent inverse cascade effects allow the development of large-scale motions 
with spatial scales of the size of the system and temporal variations much longer 
than the convective turnover time. 
These motions can be separated into the longitudinal and meridional components
namely, the differential rotation and meridional flow. 
The resulting differential rotation for simulation R1x24 is 
solar-like, accelerated (decelerated) at the equator (poles) with respect to the 
frame rotation rate $\Omega_0$, see panels Fig.~\ref{fig:mean_flows}(a,b). 
Yet, unlike the Sun where iso-rotation contours are conical,  the profile 
has contours roughly cylindrical, aligned with the rotation axes. 

The ambient state prevents the turbulent motions from penetrating into 
the stable layer below the thin overshooting region. Thus, a tachocline is 
formed at the transition between the radiative and convective zones. 
In the Sun, the tachocline is subjected to radiative 
spreading \citep{spiegel+zahn92},
and there is not yet a widespread agreement on the mechanisms that
sustain its thickness. 
In most simulations including a sub-adiabatic layer below the convection zone,
the tachocline spreads over time due to the imposed viscosity. 
In the ILES simulations presented here,
there is no measurable viscous spreading of the tachocline in the stable
layer during timescale of the simulations. This allows us to assess its
role on the formation and sustain of large-scale flows.  

The meridional circulation is represented in Fig.~\ref{fig:mean_flows}(c)
through the mean latitudinal velocity in the
meridional plane. In the northern hemisphere, the negative (positive) values
of $\mean{v}$ correspond to poleward (equatorward) motions. The graph
shows several circulation cells appearing at latitudes between 
$\pm 35^{\circ}$ with amplitudes of a few m s$^{-1}$ with the most
prominent cells corresponding to clockwise circulation.  
At higher latitudes the meridional flow 
weakly reaches the poles at the upper layers and returns at deeper layers. 

The DR of simulation R2x24, Fig.~\ref{fig:mean_flows}(d,e) still shows 
a fast equator. Nevertheless, a column of strong retrograde velocity appears 
outside the cylinder tangent to the model's tachocline.  Physically, this is 
an expected outcome if the velocity of counterclockwise convective
Busse columns is enhanced due to a large Rossby number 
\citep[see e.g.,][]{FM15}.
Inside the tangent cylinder, there is a column of rotation with roughly the
same speed as the equator, and slower poles at the highest latitudes. 
The same characteristics appear in simulation R4x24, however, increasing the
resolution further decreases the equatorial speed and enhances the acceleration
of the poles.   The profiles of meridional flow remain similar in these
simulations yet with stronger meridional flow velocities at intermediate to 
high latitudes. These enhanced meridional flows transport angular momentum 
towards the higher latitudes and explain the obtained polar acceleration. 

Even though a solution independent of the grid resolution is not observed,  
and the highest resolution simulation is the farthest from 
the solar-like rotation, with the data at hand, it is possible 
to explore the redistribution of angular momentum  as a function of the 
resolution to get a better understanding of the processes that drive 
and sustain the mean-flow profiles. 

The mean angular momentum, 
${\cal L} = \varpi {\rho}_r \mean{u}$, where $\varpi = r \sin\theta$ is the
lever arm and $\mean{u}$ is the time and longitudinal average of $u$,  
evolves  according to:
\begin{equation}
  \frac{\partial {\cal L}}{\partial t}
     =-\nabla\cdot \left( \varpi \left[ 
       \rho_r (\mean{u}+\varpi \Omega_0)\mean{\bm u}_{\rm m} + 
       \rho_r \mean{u' {\bm u}'_{\rm m}}  \right]  \right ),
       \label{eq:amb} 
\end{equation}

where, $\mean{\bm u}_{\rm m}$, and ${\bm u}'_{\rm m}$ are the mean and 
turbulent meridional ($r$ and $\theta$) components of the velocity field, respectively.
More explicitly, the terms inside the divergence are the fluxes of angular 
momentum,
\begin{eqnarray}
 \label{eq:amf1}
 {\cal F}_{r}^{\rm RS}&=&\rho_r \varpi \mean{{u}' {w}'}, \\\nonumber
 {\cal F}_{\theta}^{\rm RS}&=&\rho_r \varpi \mean{{u}' {v}'}, \\\nonumber
 {\cal F}_r^{\rm MC}&=&\rho_r \varpi (\mean{u}+\varpi \Omega_0)   \mean{w}, \\\nonumber
 {\cal F}_{\theta}^{\rm MC}&=&\rho_r \varpi (\mean{u}+\varpi \Omega_0) \mean{v},\\\nonumber
\end{eqnarray}       
which arise from the small-scale correlations of the turbulent flow, 
cf. Reynolds stresses (RS) and from
the mean profiles of DR and MC that establish at a statistically steady state.
It is expected that during this stage, the LHS of the equation vanishes; 
therefore, the four fluxes of Eq.~\ref{eq:amf1} should balance with each other.
The net radial and latitudinal angular momentum transport may be estimated by 
computing the fluxes across spherical and conical surfaces, 
respectively \citep[see][]{brun+02} as
\begin{eqnarray}
 I_r(r)=\int_0^{\pi} {\cal F}_r(r,\theta) r^2 \sin\theta d\theta \\\nonumber
 I_{\theta}(\theta)=\int_{r_b}^{r_t} {\cal F}_{\theta}(r,\theta) r \sin\theta dr,
\end{eqnarray}
where:
\begin{eqnarray}
 \label{eq.amf2}
 {\cal F}_r={\cal F}_r^{\rm MC}+{\cal F}_{r}^{\rm RS} \\\nonumber
 {\cal F}_{\theta}={\cal F}_{\theta}^{\rm MC}+{\cal F}_{\theta}^{\rm RS}.
\end{eqnarray}
Since $r$ runs from bottom to top, positive (negative) values of $I_r$ correspond
to upward (downward) angular momentum flux. Similarly, $\theta$ runs from the 
north to the south poles; thus, positive (negative) $I_{\theta}$ corresponds to 
equatorward (poleward) flux in the northern hemisphere.
\begin{figure*}%
    \begin{center}
    \includegraphics[width=0.3\textwidth]{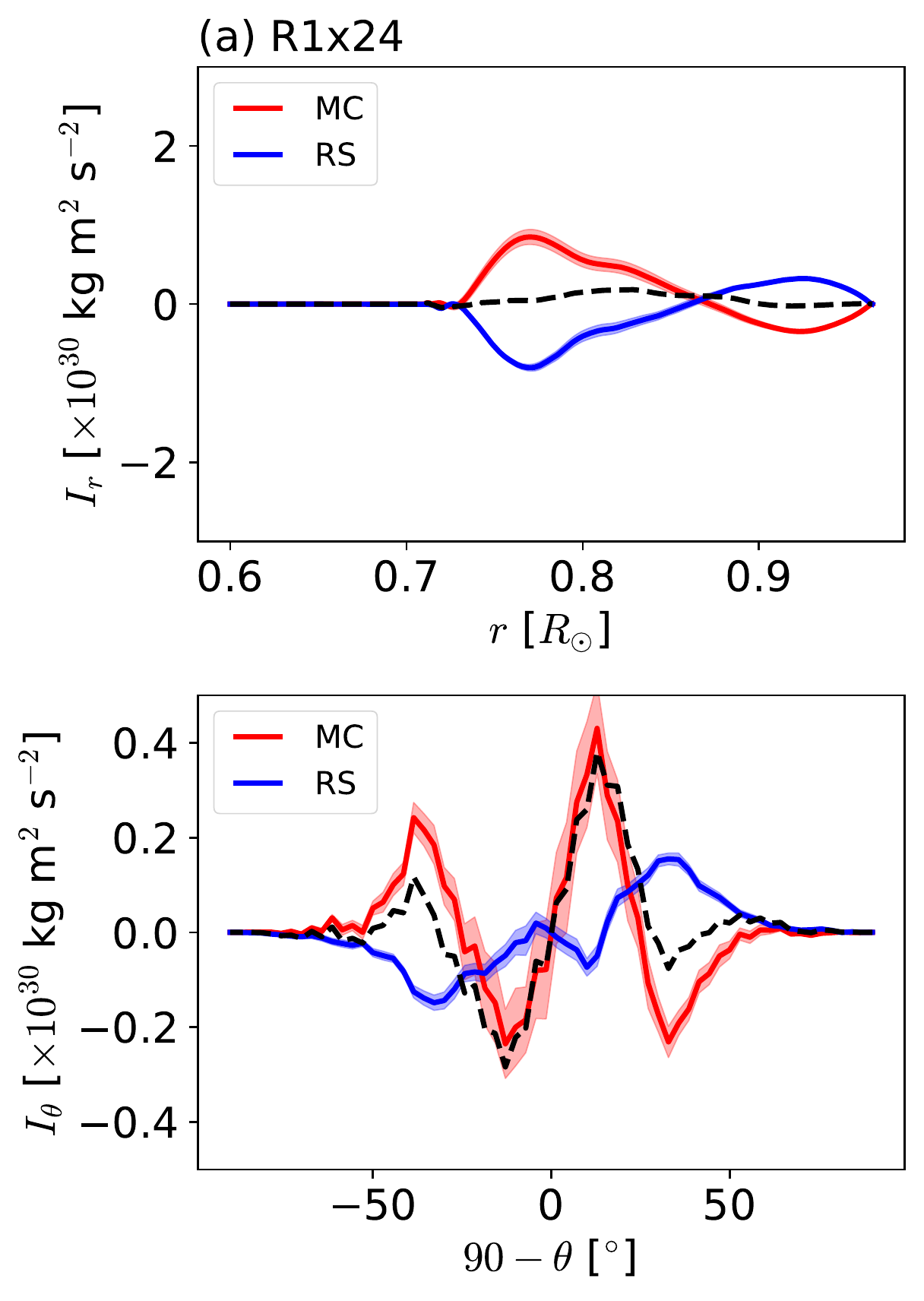} 
    \includegraphics[width=0.3\textwidth]{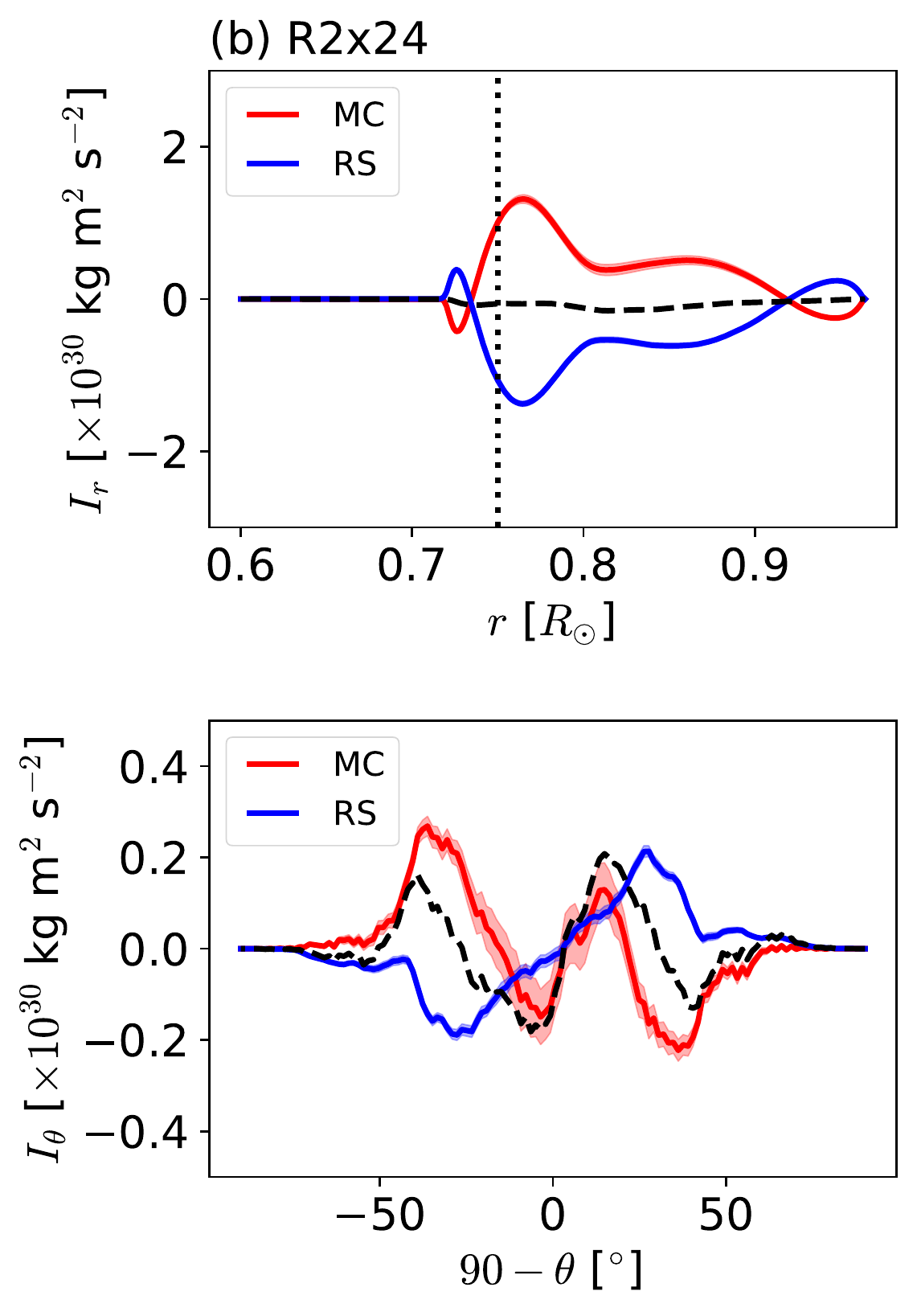} 
    \includegraphics[width=0.3\textwidth]{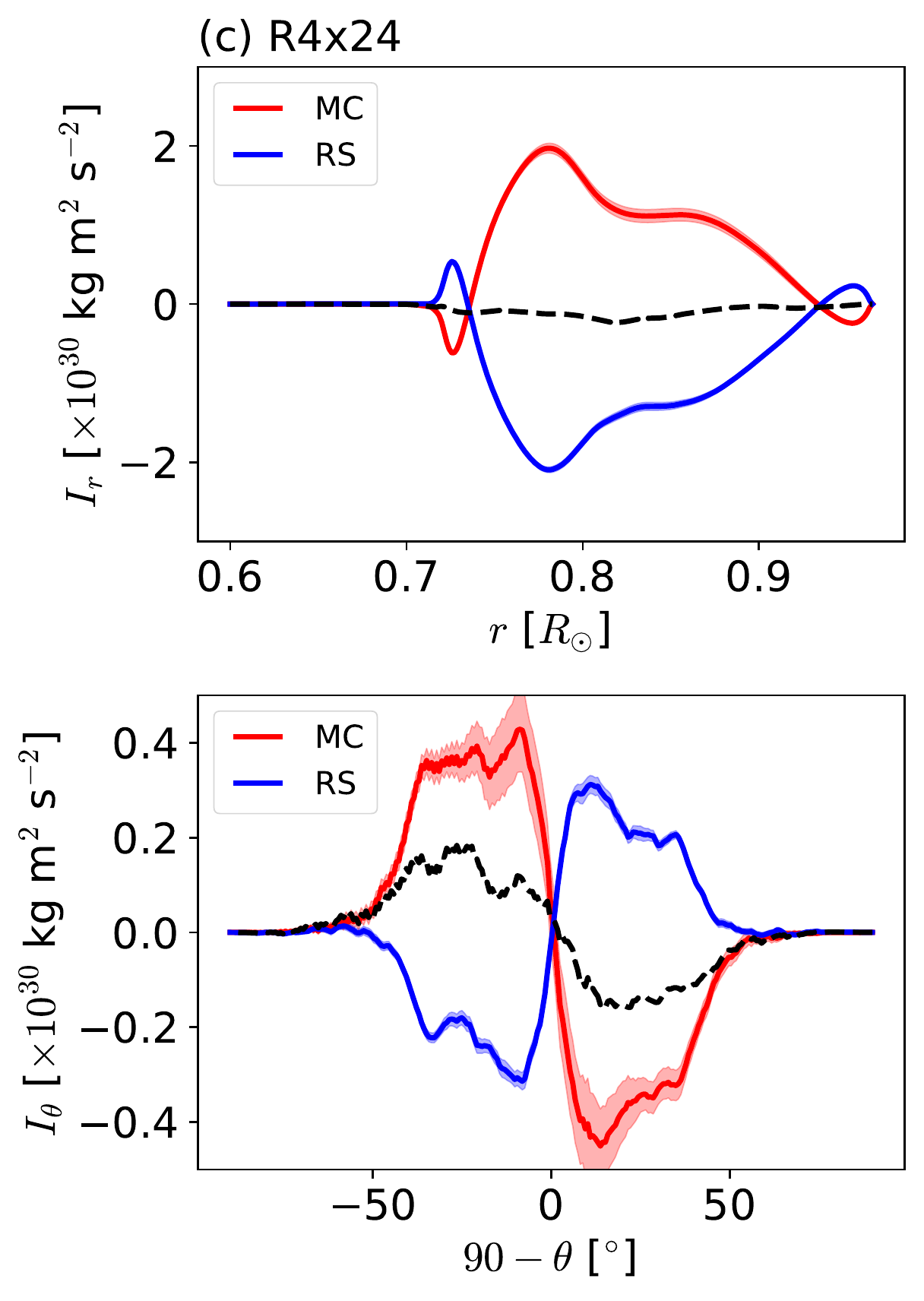} 
	    \caption{Angular momentum fluxes integrated over spherical surfaces
	    at different $r$ (upper panels) and conical surfaces at  latitudes 
	    $90-\theta$ (bottom), for the simulations (a) R1x24, (b) R2x24 and 
	    (c) R4x24. The red and blue lines correspond to the MS and RS contributions,
      the dashed black line is the sum of both contributions. }
    \label{fig:int_angmom_flux}%
    \end{center}
\end{figure*}
The fluxes of angular momentum for the
simulations R1x24, R2x24 and R4x24, integrated in radius and latitude,
are presented in the top and bottom rows
of Fig.~\ref{fig:int_angmom_flux}, panels (a) to (c), respectively. For evaluation
of the independent contributions of the RS and the MC,  
the integrals for ${\cal F}_r^{\rm MC}$ and ${\cal F}_{\theta}^{\rm MC}$ (red lines), 
and  ${\cal F}_{r}^{\rm RS}$ and ${\cal F}_{\theta}^{\rm RS}$ (blue lines) 
are presented separately. The
black dashed lines show the sum of the MC and RS contributions.

For the lowest resolution case, panel (a), the RS transports angular momentum
downwards at the bottom half of the convection zone and upwards at the upper half.
Increasing the resolution leads to an increase in the radial flux (panels b and c). 
However,  for the cases R2x24 and R4x24, the RS flux is negative, 
inward, in most of the convection zone,  with boundary 
regions of positive, upward,  flux at the bottom and top of the 
domain.  The radial RS flux is well balanced by the MC
flux, and the sum of the two fluxes is roughly consistent with zero.

The amplitude of the latitudinal fluxes also increases with the numerical
resolution.  In all the cases the RS flux pumps angular momentum towards the 
equator. Unlike the radial flux, the MC flux does not balance its turbulent
counterpart and even shows a different profile for each case. 
In simulation R1x24,  $I_{\theta}^{MC}$ has
the same sign as  $I_{\theta}^{RS}$ at lower and intermediate latitudes. Therefore,
the net transport of angular momentum is equatorward.   This explains the 
solar-like profile observed in Fig.~\ref{fig:mean_flows}(a, b). The black
dashed line in the bottom panel of Fig.~\ref{fig:int_angmom_flux}(a) clearly
shows that the two fluxes do not balance each other.  In these hydrodynamic
simulations the only term missing in Eq.~(\ref{eq:amb}) is the viscous flux. While 
in the Sun this term must be irrelevant, in the low-resolution case, it appears
in the form of effective viscosity.  This numerical contribution results in
an angular momentum flux with values compatible with the RS and MC fluxes. 

In simulations R2x24 and R4x24, the balance between RS and MC fluxes is
better, as evidenced by the smaller values reached by the black dashed line
relative to the values of $I_{\theta}^{RS}$ and $I_{\theta}^{RS}$. 
Yet, the balance is not perfect 
(bottom row of panels b and c). The MC flux assumes
the opposite sign of RS and advects angular momentum towards the poles.  
Notice that in the three cases the resolution increases only in the horizontal
direction.  The results of \citet{nogueira+22} with the EULAG-MHD code in
2D Cartesian simulations indicate that
the effective viscosity scales as $N^{-2.7}$, with $N$ being the
horizontal resolution.  Thus, changing $N_{\phi}$ and $N_{\theta}$ by
a factor of two and four  results in a change in the effective viscosity
with a factor of $\sim 7$ and $\sim 300$, respectively. Although in the spherical
coordinates the scaling of the effective viscosity with the resolution may be
different, it is not unrealistic to expect changes by a factor larger than $10$.
It is good to have in mind, however, that this viscosity is
non-linear and non-homogeneous, being larger in regions where the flow has
steeper variations.  
The bottom row of Fig.~\ref{fig:int_angmom_flux}(c) demonstrates 
the weaker influence of the effective viscosity on the angular momentum dynamics. 
In this case $I_{\theta}^{MC}$ has almost twice the amplitude of  $I_{\theta}^{RS}$.
Therefore, the sum of the two has also half of the amplitude of $I_{\theta}^{MC}$ and
its contribution aims to settle the balance. The profiles of $I_{\theta}^{MC}$  
explain the transition of the DR from the accelerated
equator to accelerated poles.  It is worth mentioning here that
the profiles in Fig.~\ref{fig:int_angmom_flux}(c) are compatible with those of
\citet{miesch+08}, with the exception of an enhanced MC latitudinal flux in 
case R4x24.  Nevertheless, unlike the differential rotation profile observed in
Fig.~\ref{fig:mean_flows}(g, h), they obtain a solar-like rotation.  There are
several differences between the two modeling approaches. The more 
relevant ones are perhaps the different values of the Prandtl number 
($0.25$ in their case and $\sim1$ in this work) and the initial condition of the 
simulation as well as the total evolution time.  They enforce a solar-like profile and then 
evolve the simulation for $\sim3$ years, whereas the models presented here start 
from random thermal fluctuations and evolve the system for $20$ years. 

The large-scale flows observed in Fig.~\ref{fig:mean_flows} are 
sustained by the fluctuations around average values of the angular momentum
fluxes.  The shadow involving the profiles presented in Fig.~\ref{fig:int_angmom_flux} 
depict the standard error, $\sigma_e = \sigma/\sqrt(n)$, where $\sigma$
is the variance and $n$ the number of samples in the average.  It is clear
from the figure that the MC flux has larger deviations from the mean profile and
that these deviations increase for higher resolutions. This result is not
surprising given the smaller viscous friction that the flow experiences
in the higher resolution cases. The fluctuations are sustained by 
turbulent convection in turn driven by the buoyancy force. Therefore, 
Eq.~(\ref{eq:amb}) does not entirely describes the sustainability of the 
mean-flows. 

\begin{figure}%
    \begin{center}
    \includegraphics[width=\columnwidth]{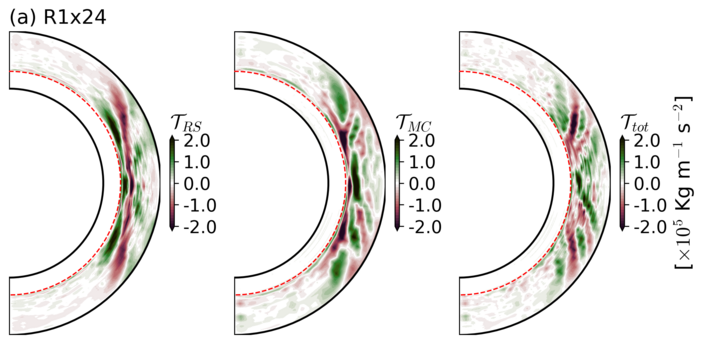} 
    \includegraphics[width=\columnwidth]{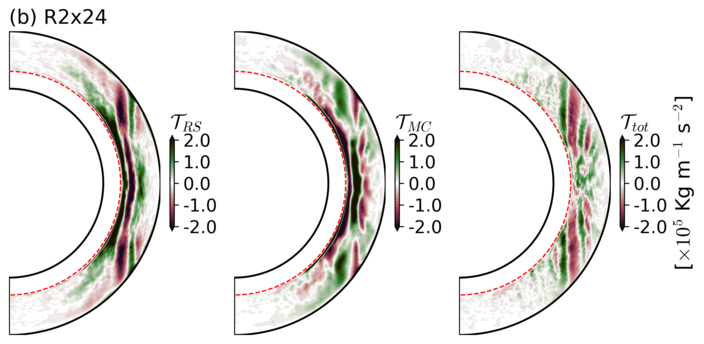}
    \includegraphics[width=\columnwidth]{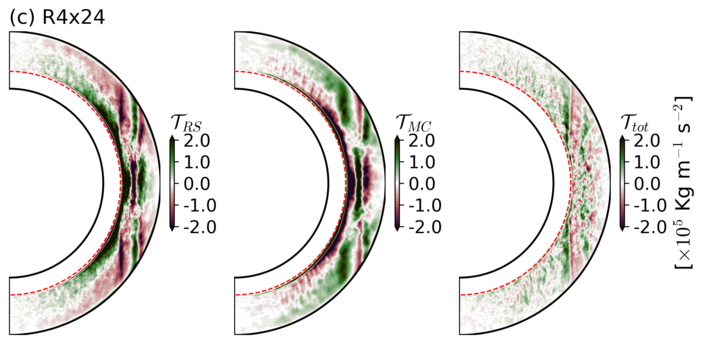}
	    \caption{Axial torques due to the RS (left column), the MC (middle), and
	    the sum of the RS and MC (right) torques presented in the meridional plane
      for the simulations R1x12 (top row), R2x24 (middle) and R4x24 (bottom).}
    \label{fig:axial_torque}%
    \end{center}
\end{figure}

Before providing an overview on the sustainment of the DR and MC, it
is illustrative to compute the divergence of the fluxes on the RHS of 
Eq.~(\ref{eq:amb}). Including the minus sign in front, the results are the axial 
torques; 
\begin{eqnarray}
{\cal T}^{RS} &=& -\frac{1}{r^2}\frac{\partial (r^2 F_r^{RS})}{\partial r} 
               - \frac{1}{r \sin\theta}\frac{\partial (F_{\theta}^{RS} 
               \sin\theta)}{\partial \theta} \;, \\
{\cal T}^{MC} &=& -\frac{1}{r^2}\frac{\partial (r^2 F_r^{MC})}{\partial r} 
               - \frac{1}{r \sin\theta}\frac{\partial (F_{\theta}^{MC} 
               \sin\theta)}{\partial \theta} \;,
\end{eqnarray}
due to the RS and the MC, respectively.  These quantities are presented in
Fig.~\ref{fig:axial_torque}.  
On average, in the steady-state, the RS produce a torque that changes sign with
radius and latitude. The profiles from simulations R1x24 to R4x24 are similar,
although, for the higher resolution cases, the torque becomes stronger, 
better aligned with
the axis outside the tangent cylinder, and seems to reach higher
latitudes. For balancing these torques, the meridional motions form closed loops
whose circulation has axial torques in the opposite direction to the RS. 
This process is called gyroscopic pumping. It is sustained from deviations of the
thermal wind balance as will be seen in the following section.  The direction 
of the meridional flows may be identified by writing Eq.~(\ref{eq:amb}), 
with $\partial {\cal L}/\partial t =0$, as,
\begin{equation}
  \rho_r  \mean{{\bm u}}_m \cdot \nabla (\varpi^2 \Omega)  = -\nabla \cdot 
  \left(\rho_r \varpi \mean{{\bm u}_m^{\prime} u^{\prime}}\right) = -{\cal T}^{RS},
  \label{eq:gyro}
\end{equation}
and with the known profiles of $\Omega$ and ${\cal T}^{RS}$ 
(Figs.~\ref{fig:mean_flows} and \ref{fig:axial_torque}). 
As a simple example, let's consider the simulation R4x24, with constant $\theta$ 
at the equator, equation~(\ref{eq:gyro}) becomes
\begin{equation}
  \rho_r \mean{w}_{eq}(r) \frac{\partial}{\partial r}(r^2 \Omega_{eq}(r) ) = 
  - \frac{1}{r^2}
  \frac{\partial}{\partial r}(r^2 F_{r,eq}^{RS}(r)).
\end{equation}
The radial derivative at the LHS is positive for all $r$, whereas the 
radial derivative at the RHS is negative for $0.72\Rs \lesssim r \lesssim 0.77$, 
positive for $0.77\Rs \lesssim r \lesssim 0.82$ and negative again for 
$0.82\Rs \lesssim r \lesssim 0.9$, 
from where it is positive, see the leftmost panel of Fig.~\ref{fig:axial_torque}(c).  
Thus, because of the minus sign in the RHS, the 
profile of $ \mean{w}_{eq}(r)$ has to have the opposite sign to 
$\frac{\partial}{\partial r}(r^2 F_{r,eq}^{RS}(r))$. We have numerically
verified that this relation is obeyed for all values of $r$. 

In general, Fig.~\ref{fig:axial_torque} shows that the MC torque roughly has 
the same profile than the RS torque yet with the opposite sign. 
The third column of the figure depicts ${\cal T}_{tot} = {\cal T}^{RS} + {\cal T}^{MC}$.
It shows that the balance is not perfect, suggesting the contribution 
of axial torques by the effective viscosity.  
With the increase of the resolution, panels (a) to (c), 
the values of the axial torques become higher, the balance seems
to improve, and ${\cal T}_{tot}$ is
evidently less coherent. Both changes are a consequence of the diminished 
viscous resistance to the turbulent motions. 

\subsubsection{Differential temperature and thermal-wind balance}
\label{sec:diff_temp}

Both quantities on the LHS of Eq.~(\ref{eq:gyro}) are sustained from the departures of 
the equilibrium state through fluctuations driven by convection, which in turn is 
sustained by the buoyancy force. An equation including this contribution may 
be obtained by computing the vorticity by taking the curl of 
Eq.~(\ref{eq_momentum_nc}). The longitudinal component of the  vorticity  
contains various terms that sustain the meridional balance 
\citep[see e.g.,][for derivation and discussion]{MH11,Passos+17}.  
The inertial, ${\cal I}$, and the baroclinic, ${\cal B}$, terms are the most 
relevant \citep{kitchatinov_2014} leading to the thermal wind balance (TWB) 
equation,
\begin{equation}
  {\cal I} = \varpi \frac{\partial \Omega^2}{\partial z} = {\cal B} + {\cal D} =  
  \frac{g}{\Theta_r} \frac{\partial \Theta^{\prime}}{\partial \theta} + {\cal D},
  \label{eq:twb}
\end{equation}
where $z$ is the vertical axis in cylindrical coordinates, such that
$\partial_z = \cos\theta \partial_r - r^{-1}\sin\theta\partial_{\theta}$, and 
${\cal D}$ incorporates all other forces in the meridional plane, including the 
contribution of the diagonal as well as the meridional components of the Reynolds 
stress tensor. 

\begin{figure*}%
    \begin{center}
    \includegraphics[width=0.3\textwidth]{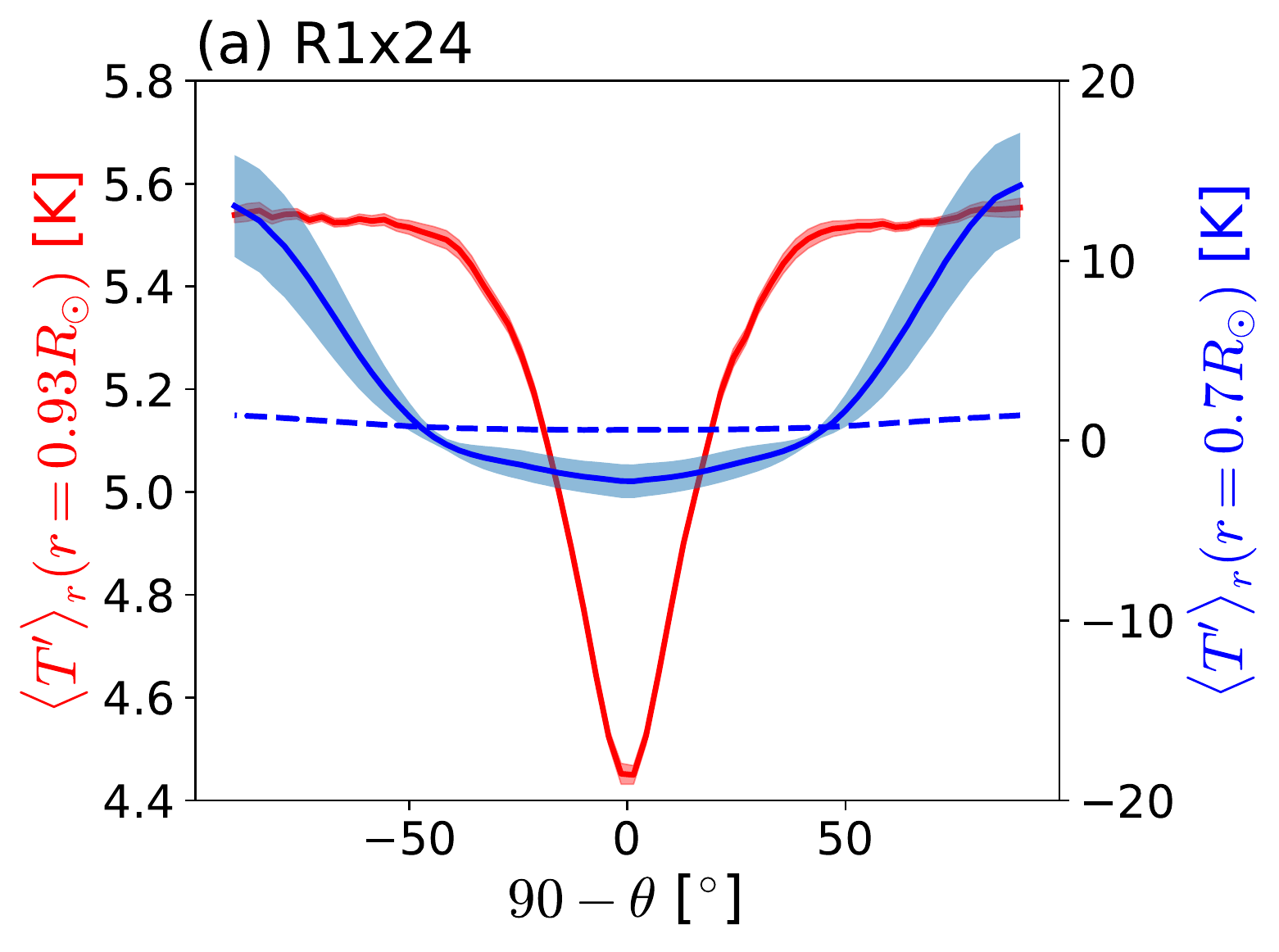} 
    \includegraphics[width=0.3\textwidth]{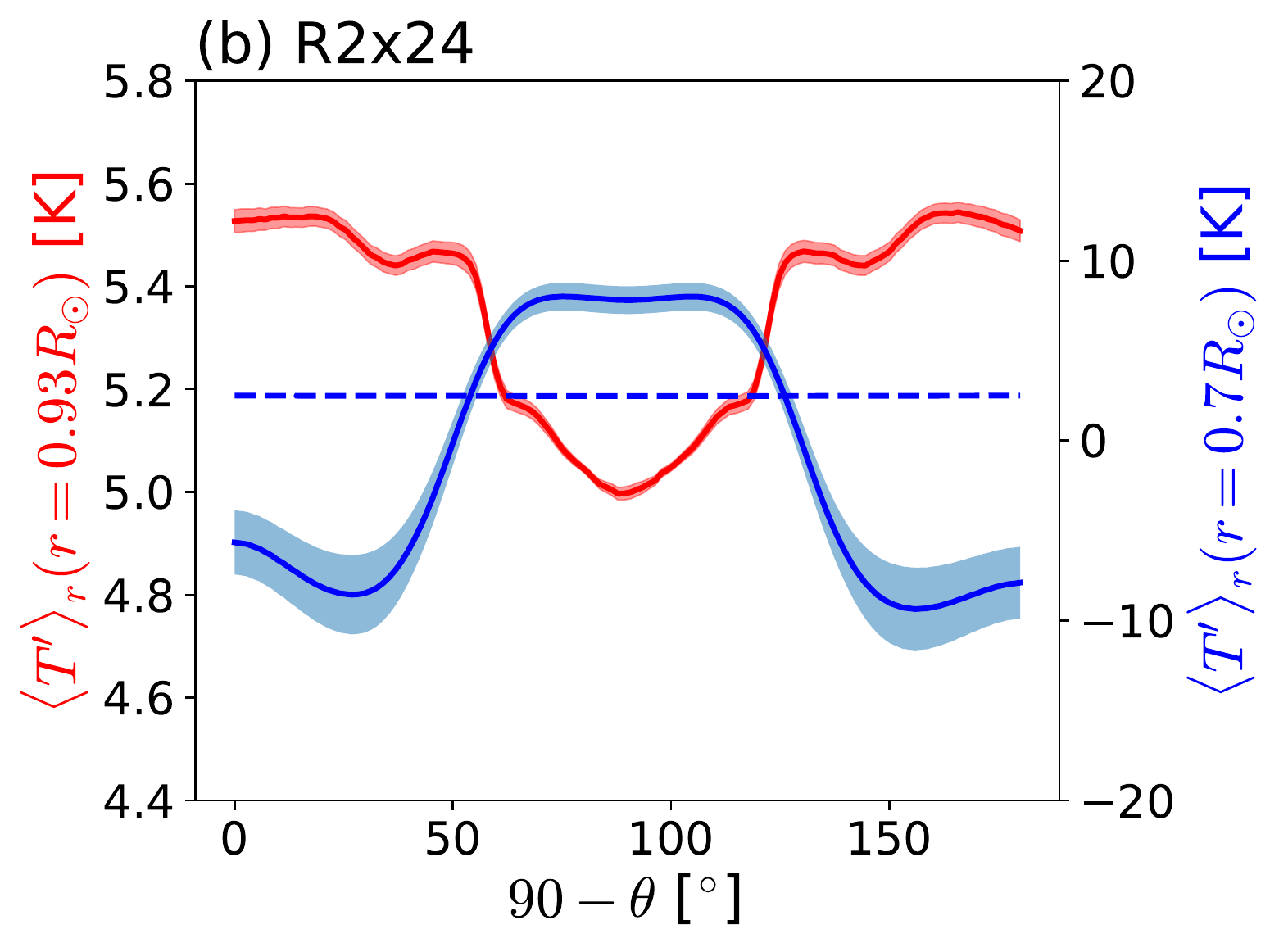} 
    \includegraphics[width=0.3\textwidth]{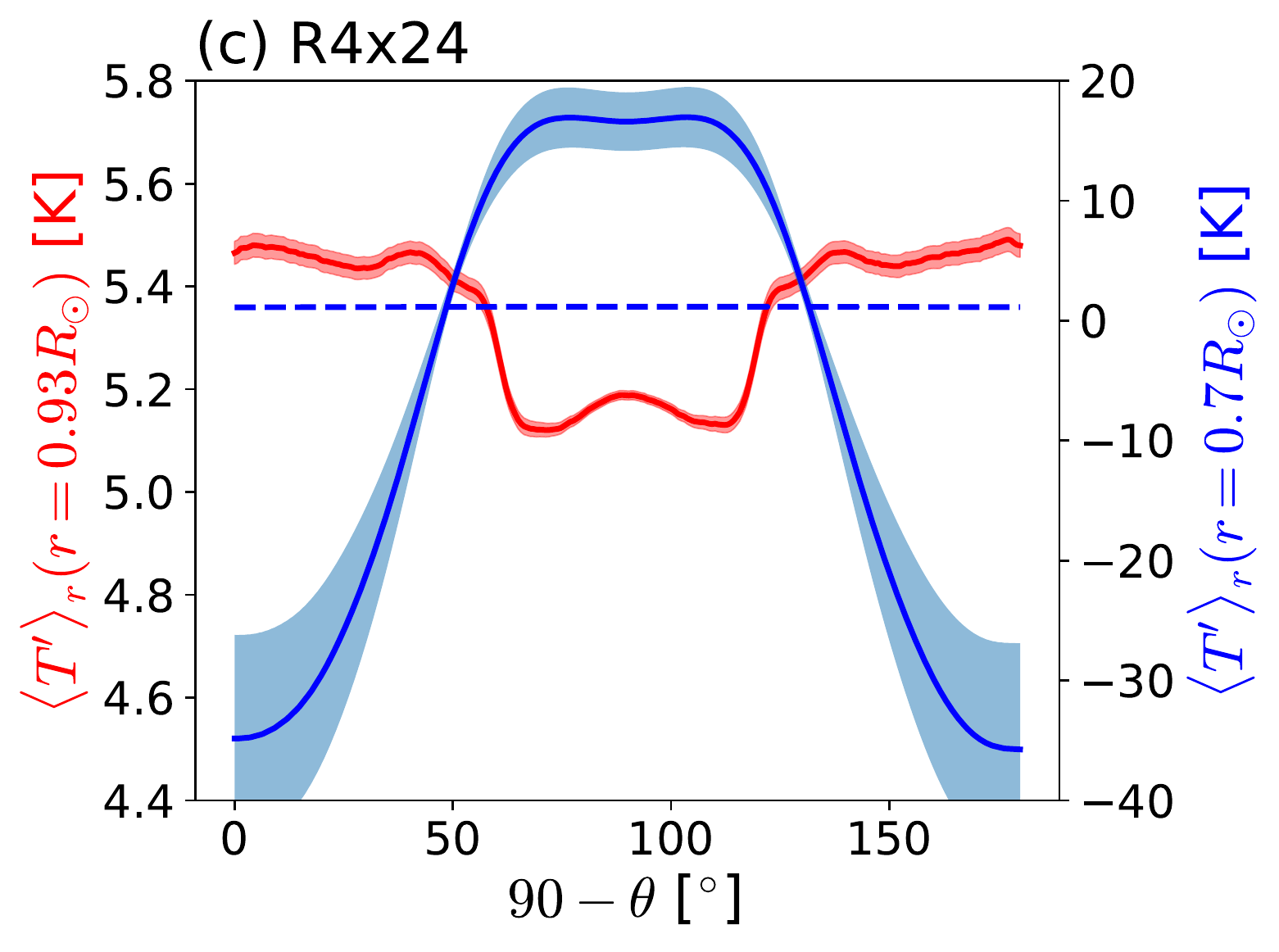} 
	    \caption{Latitudinal variations of temperature perturbations: 
      the red and blue lines correspond to
	    profiles of temperature perturbations at the surface and the bottom of 
	    the convection zone for the models (a) R1x24, (b) R2x24, and (c) R4x24.
	    The contrast at the two radial levels is different; therefore two different 
      scales at the left and right $Y$ axis are considered. The dashed line
	    shows a third radial level at $r=0.66$, where perturbations are not
      expected. The $\brac{}_r$ indicates radial average centered in $r=0.71\Rs$
      and $r=0.93\Rs$ in a radial layer of thickness $0.04\Rs$.}
    \label{fig:diff_t}%
    \end{center}
  \end{figure*}

Departures from the TWB due to fluctuations of the 
LHS term induce meridional motions via the gyroscopic pumping seen from 
Eq.~(\ref{eq:gyro}).
Similarly, latitudinal differential temperatures result in meridional flows 
driven by baroclinicity.  Equations (\ref{eq:gyro}) and (\ref{eq:twb}) 
are obviously coupled.  As mentioned above, meridional motions are necessary to 
balance the angular momentum, these motions are generated by deviations from the 
TWB. Fig.~\ref{fig:diff_t} shows the temperature
fluctuations as a function of latitude.  From the figure, it is possible to
determine whether the gyroscopic pumping or the
baroclinic force drives the meridional flow.
For better visualization of the latitudinal differential temperature,
two different scales are considered at the left and right
of the panels (a) to (c), corresponding to simulations R1x24 to R4x24, 
respectively. The scale at the left (right) $Y$-axis shows $T^{\prime}$ at the 
top (bottom) of the convective layer. 
Comparing  the blue and red curves indicates that 
the latitudinal contrast is small (less than one degree) at the top of the domain, 
yet it reaches tens of Kelvins at the bottom of the CZ. Furthermore, 
whereas at the top of the domain the equator is colder than the poles for all 
cases, below the CZ the equator is colder only for simulation R1x24, and large 
temperature contrast between a warmer equator and colder poles is found 
for simulation R4x24. The shadows
show the deviation from the mean. 

Given the profiles of Fig.~\ref{fig:diff_t}, baroclinicity should result in 
motions with  prominent clockwise circulation for the case R1x24, 
and prominent counterclockwise circulation for cases R2x24 and
R4x24. This is clearly not observed in the rightmost panels of 
Fig.~\ref{fig:mean_flows},  except perhaps at intermediate to higher 
latitudes in case R4x24. The main 
source of the meridional motions observed outside the tangent cylinder is
likely the departure from thermal wind balance about the inertia term, ${\cal I}$,
which corresponds to the gyroscopic pumping. 

\begin{figure}%
    \begin{center}
    \includegraphics[width=\columnwidth]{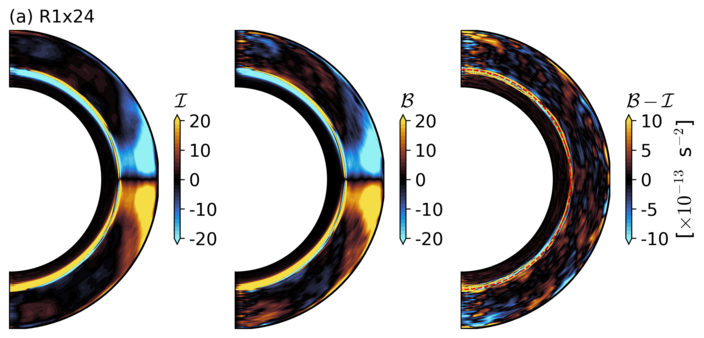}
    \includegraphics[width=\columnwidth]{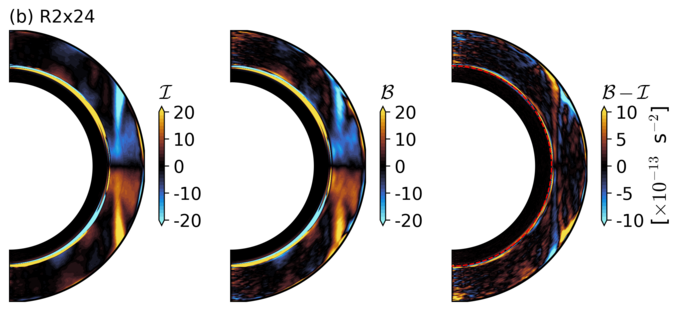}
    \includegraphics[width=\columnwidth]{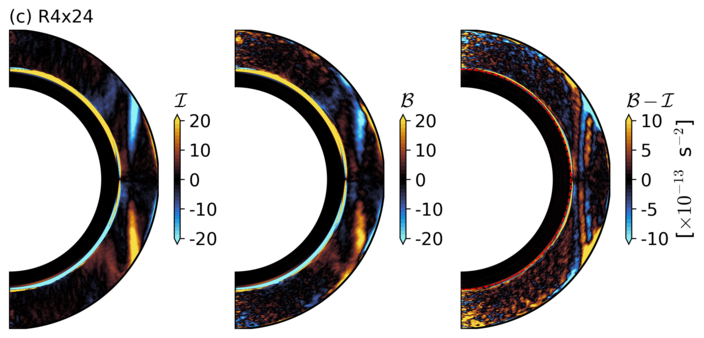}
            \caption{Inertia, ${\cal I}$ (left),  and baroclinic, ${\cal B}$ (middle) 
            terms of the TWB, Eq.~\ref{eq:twb}. The right column shows the residual
            ${\cal B} - {\cal I}$.}
    \label{fig:thermal_wind}%
    \end{center}
\end{figure}

The contribution of the meridional forces to the generation of meridional 
circulation can be better understood through the comparison of the two 
most important terms of the meridional force balance equation, namely the inertial,
${\cal I}$, and the baroclinic, ${\cal B}$, terms. If the two terms cancel 
to each other, the system is in TWB. Whenever the balance is violated, either 
transiently or after temporal and longitudinal averages, vorticity in the meridional 
plane, i.e.,  meridional circulation, is induced. 
Fig.~\ref{fig:thermal_wind} shows contours of ${\cal I}$ (first column), 
and ${\cal B}$ (second column) in the meridional plane. The third column  
shows the departure of TWB via the residual ${\cal B}-{\cal I}$. 

The first thing to notice in Fig.~\ref{fig:thermal_wind} is the qualitative and
quantitative similarity of the meridional profiles ${\cal I}$ and ${\cal B}$. 
However, the third column shows that the balance is not perfect. Note that the
scale of the residual is half of the inertial and baroclinic terms. Thus, 
the deviation from TWB is smaller yet considerable. As expected, this deviation
is more prominent outside the tangent cylinder, where
the meridional circulation cells are observed. Not surprisingly, 
it increases with the numerical resolution providing enhanced meridional circulation 
(note that the opposite happens with the 
axial torques, Fig.~\ref{fig:axial_torque}, i.e.,  the residual decreases with 
the increase of resolution).  

The departures from the TWB by meridional
drivers such as the meridional Reynolds stress component sustain a meridional
flow which is stronger at higher resolution because the numerical viscosity
offers less and less resistance to these motions.  The meridional motions
compensate for the angular momentum transported by the Reynold stresses. In the
low resolution case, where viscosity is high, the angular momentum balance
due to MC is insufficient and the positive (negative) values of $I_{\theta}^{RS}$ 
in the northern (southern) hemisphere efficiently accelerate the equator.  
In the high-resolution case, the MC motions develop with less friction and
better compensate for the angular momentum fluxes (see $I_{\theta}^{MC}$ in 
Fig~\ref{fig:int_angmom_flux}). As a matter of fact, they are sufficiently 
strong to accelerate the poles. In the Sun, the molecular viscous resistance is 
insignificant. Thus, there is always the magnetic field, which may provide
the large or small-scale Maxwell stresses necessary to allow 
an effective equatorial transport of angular momentum resulting in solar-like 
rotation. Yet it is not the only possibility to solve this issue.

In mean-field models of differential rotation 
\citep[see][for a review]{kitchatinov_2014}, the TWB is almost exact in
the bulk of the convection zone.  And departures are obtained at the
boundary layers in such a way that a strong poleward meridional flow develops
at the near-surface shear layer, and the equatorial return flow occurs at 
tachocline  levels.  The theory for the formation of the NSSL developed by 
\citet{MH11} is in agreement with this view, suggesting that, in this layer, the 
inertia term is balanced by turbulent stresses rather than by the baroclinic 
force. It is worth remarking that in the simulations presented here 
the NSSL is not considered, and these boundary effects may be a relevant
missing element.  Nonetheless, baroclinicity is not sufficient to balance the
inertial meridional force as presented above. It is possible 
that the turbulent motions developed in most of the current 
global simulations are not a reliable representation of the high $\Re$, 
low $\Pr$ turbulence  occurring inside the Sun. 

Another remarkable point from Fig.~\ref{fig:thermal_wind} is 
that the tachocline does not seem to act as a boundary layer generating strong
departures of the TWB as suggested in mean-field models.  
It has usually been assumed that latitudinal gradients of temperature may
be responsible for the deviation from the cylindrical towards the 
conical iso-rotation contours observed in the Sun.  This seems to work 
in global simulations that do not include the tachocline where a latitudinal 
differential temperature is imposed as a boundary condition \citep{miesch+06}.  
The simulations presented in 
this work, including the tachocline, demonstrate otherwise. Consider,  
for instance, the low-resolution simulation, R1x24. The slower north pole and 
fast equator make the variation of $\mean{\Omega}$ along the direction of
the rotation axes ($z$) most negative at the
poles, than at $\sim 40^{\circ}$, where the convection zone rotates at
the same speed than the stable layer. This is compensated by a negative
gradient of $\Theta^{\prime}$, and consequently in $T^{\prime}$.  
Conversely, in case R4x24, $\mean{\Omega}$ goes from a slower stable
layer towards a faster north pole, generating a positive signal for 
the term ${\cal I}$. This is compensated by a positive temperature gradient 
of  $\sim 40$ K which is also evident in ${\cal B}$.
Therefore, it is possible to conclude that, in our simulations, the 
differential temperature is a reaction of the baroclinic force to the 
gradient of the angular velocity the $z$ direction. Once the balance is 
established, the outcome are roughly cylindrical iso-rotation contours 
independently of the rotation being faster at the equator or at the poles.

\section{Conclusions}
\label{sec:conclusions}

We have performed HD convection simulations with the EULAG-MHD code in a spherical 
shell whose thermodynamic stratification resembles the upper part of the solar 
radiation layer and the convection zone up to $r=0.96\Rs$. We have considered cases 
without and with rotation. The thermal driving, considered here through 
an ambient state, results in an enthalpy flux corresponding to $\sim 0.3L_{\odot}$ 
for the non-rotating cases and $\sim 0.2L_{\odot}$ for the rotating simulations. 
Keeping fixed all the parameters in the governing
equations, we explore
how progressively increasing the numerical resolution affects the simulation results.
We solve the equations in their inviscid form, therefore, the only dissipation
of momentum and heat is delegated to the truncation terms of the numerical
method. Arguably, in this form we can achieve larger values of the 
effective Reynolds number with less computational resources.  
It is expected that at sufficiently high Re, the contribution of viscosity
to the transport of linear and angular momentum should be negligible and
the dynamics of the system is defined by the properties of the turbulent motions.
Our main goals are to identify the properties of the fluid as a function of the
resolution and evaluate the challenges of achieving grid independent results
for this complicated problem.

For the non-rotating cases our findings can be summarized as follows:
\begin{itemize}
  \item
The low-order velocity moments and turbulent spectra of the non-rotating simulations
do not show significant changes between the highest resolution cases R4 and R8, 
indicating that the ILES convergence might be achieved at a resolution not much 
higher than R8.
  \item
The ambient state considered in the model captures the sharp transition
in the Brunt-Väisälä frequency.  High resolution in the radial direction
is needed for this stiffness
to be captured by the velocity field, especially its horizontal components.  
The gradient of $U_{\rm rms}$ between stable and unstable layers increases 
with $N_r$ but seems to saturate at $N_r = 256$ grid points 
(this value corresponds to a radial grid size, $dz=982$ km). 
This resolution
also appears sufficient to capture the 4.5 density scale heights considered in
the model (3.7 only in the convection zone).  
\item 
The convective motions resulting from cases R4 and R8  are characterized
by anisotropic convection with most of the energy in the divergent component
of the horizontal flow at the top of the domain. 
Because of the thickness of the convection zone, the power spectra peaks
at larger scales, 1000-1400 Mm ($\ell \sim 3-4$),  in good agreement with 
similar simulations performed at a much higher resolution \citep{hotta+14,hotta+19}.
There is also agreement with some characteristics of supergranulation
obtained by  \citet{rincon+17}. These motions, of course, have maximum energy 
at much smaller scales presumably because are driven by thermal perturbations injected
in a thin boundary layer.
\item
In case R1x, we increased the resolution of the simulation in the $r$ direction
and kept $N_{\phi}$ and $N_{\theta}$  equal to case R1. Despite the lack of 
detailed small-scale structures, the flow properties of this case correlate well 
with those obtained for cases R4 and R8. For this reason, we consider this 
resolution our starting point for the rotating simulations. 
\end{itemize}

More important for the solar dynamics is the case of rotating convection. 
We performed three simulations where the radial grid size is kept constant, and
only the horizontal resolution is increased.  
These cases appear in Table~\ref{table:1}
as R1x24, R2x24 and R4x24.
The rotation period of these simulations
is 24 days. It is slightly shorter than the solar period, yet it results in 
solar-like differential rotation in the lower resolution experiment, R1x24.  
The main results of these simulations are summarized below:

\begin{itemize}
  \item
The perturbations of $\Theta$, and more importantly, the luminosity 
carried by the enthalpy flux,
are roughly the same for all simulations. Nevertheless, the amplitude
of the velocity components increases with the resolution. This is more evident
in the RMS profile of the radial component, $w_{\rm rms}$. The difference
seems to be larger between R4x24 and R2x24, than between R2x24 and R1x24.
This suggests that for simulations accounting for the rotation, grid independent 
solutions are hard to achieve.
\item
The turbulent kinetic spectra shows strongly anisotropic motions 
near the top of the convection zone. The anisotropy decreases at the middle 
and bottom of the convection zone. As expected, rotation imprints vorticity in the
flow. Thus, the energy of the toroidal part of the horizontal flows becomes 
dominant. Interestingly, the scale with maximal energy in the horizontal motions 
appears independent of the grid size. On the other hand, the energy in the radial 
component increases with the resolution, and the peak of the spectrum shifts 
towards larger harmonic degrees, i.e, the smaller scales become more energetic. 
This seems to be a robust result when comparing our findings with those of
the high resolution simulation of \citet{miesch+08}.
\item
The change in the resulting mean-flows as a function of grid resolution is dramatic.
The  output values of the Rossby number go from $\sim 0.56$ to $\sim 0.65$, 
and in this range, we observe a transition from solar-like to anti-solar differential 
rotation.  The meridional circulation shows multiple cells aligned to the cylinder tangent
to the tachocline.  The meridional speed increases with the resolution, 
as it does the high latitudinal poleward flow.
\item
The physics behind the differential rotation transition is understood through 
the integrated angular
momentum balance and thermal wind balance (TWB). It may be summarized as follows. 
At low resolution,
the total viscosity is high; this offers strong resistance to the meridional
motions driven by transient departures of the TWB.  In this case, 
the angular momentum 
balance in the latitudinal direction is dominated by the viscosity and the 
Reynolds stresses which pump angular momentum towards the equator.  
With the increase of the 
resolution, the effective viscosity might decrease sharply, consistent with
the 2D results of \citep{nogueira+22}. Thus, 
there is a better angular momentum balance between 
the Reynolds stresses and the meridional circulation, arguably the most 
relevant processes to account for in HD simulations.
This angular momentum equilibrium favors polar acceleration because 
there is a steady departure in the meridional force balance and strong poleward
meridional flows carry angular momentum with small friction.
\item
In the simulations presented here, the latitudinal variations of temperature at the 
base of the convection zone are a consequence of TWB and not 
a results of departure from TWB as commonly
assumed (see the concluding paragraph in \S~\ref{sec:diff_temp}).  Thus, although 
there are strong temperature gradients between the equator and pole, 
the contours of iso-rotation are roughly cylindrical.
\end{itemize}

The results presented here correspond to an enthalpy flux carrying  
only one-fourth of the solar luminosity, and the rotation period is 
compatible to the sideral rotation period of the Sun. 
Extrapolating the observed trend to solar values would
result in a larger discrepancy with the solar differential rotation.  Also,
the energy contained in scales corresponding to $\ell \sim 30-40$ would 
be higher than what it is found in recent solar observations \citep{proxauf_phdT},
suggesting that increasing the resolution does not lead to a 
solution of the convective conundrum.
These problems, of course, are not specific to the EULAG-MHD simulations but 
are present in most global convection models. The results of the recent simulations 
of \citet{hotta+21} suggest that the small-scale dynamo might be a solution. 
This field  may act as friction that quenches the velocity 
and contributes to the angular momentum balance, leading to solar-like rotation.   
Other MHD simulations 
indicate that the large-scale magnetic field may contribute in the same
direction \citep{fan+14,KKBOP15,guerrero+16b}.  To this date, there are no
simulations where both, small- and large-scale contributions of the magnetic 
field, the usual suspect, are considered.  

Nonetheless, the entire problem may reside in the fact that the convection
driven in the global simulations is not an accurate representation of
the turbulent motions in the solar interior.  The reason for this could be the
enormous differences in the parameter regime between the Sun and the 
simulations. It has also been suggested that the deep solar convection may 
depend on  the physics occurring in the uppermost
50Mm (or less) of the Sun \citep{spruit97},  which are barely considered 
in global simulations.  
In this boundary layer, the effective cooling provided by hydrogen ionization 
destabilizes the plasma generating cold downward plumes.  Thus, the turbulent
motions in the deep solar convection zone may be driven by this penetrative 
entropy rain \citep{brandenburg16}.
In this sense, deep solar convection can be seen as a non-local process
in a buoyantly neutral layer. The numerical experiments performed by \citet{cossette+16}
demonstrated this concept in a Cartesian box. Their results show that the 
dominant convective scales depend on the thickness of the boundary layer.
Concurrently,  the fast convective motions in this layer are not constrained 
by the solar rotation and
produce a negative shear which, in turn, accelerates meridional motions
producing the observed surface latitudinal velocity \citep{MH11}.  
Recent simulations of \cite{kitiashvili+22} showed that turbulent convection in the upper
part of the NSSL can generate the radial differential rotation and 
meridional circulation. However,  it is still uncertain
how the turbulent correlations generated by non-local convection can 
sustain the solar differential rotation below the NSSL. Thus, it is worth
exploring this possibility and its consequences for the mean-flows in global 
simulations.   The pursuit of numerical convergence is also worth more
computational efforts. A database of simulations with varying resolutions
is necessary for studying the changes in the turbulence properties
at different Rayleigh and Reynolds numbers. It can also be used for the
experimentation of sub-grid scale methods that allow realistic simulations at a
lower computational cost. These goals will be pursued in future works. 

\begin{acknowledgments}
We thank Bonnie Zaire for her comments on the manuscript.
%We thank the anonymous referee for the constructive comments and
%suggestions that have improved the quality of the paper.
This work was partly funded by NASA grants NNX14AB70G,  80NSSC20K0602, and 
80NSSC20K1320.  NCAR is sponsored by the National Science Foundation.
The simulations were performed in the NASA supercomputer Pleiades  
\end{acknowledgments}

%%%%%%%%%%%%%%%%%%%%%%%%%%%%%%%%%%%%%%%%%%%%%%%%%%%%%%%%%%%%%%%%%%%%%%%%%%%%%
%%%%%%%%%%%%%%%%%%%%%%%%%%%%%%%%%%%%%%%%%%%%%%%%%%%%%%%%%%%%%%%%%%%%%%%%%%%%%

\bibliography{bib}{}
\bibliographystyle{aasjournal}

\end{document}